\documentclass[11pt]{article}
\usepackage{epsfig}
\usepackage{amsmath}
\usepackage{amsfonts}
\usepackage{amssymb}

\numberwithin{equation}{section}

\title{The symplectic Kadomtsev-Petviashvili hierarchy and rational solutions
of Painlev\'e VI}
\author{Henrik Aratyn\\
\\
Department of Physics,\\
University of Illinois at Chicago,\\
845 W. Taylor St.,\\
Chicago, IL 60607-7059\\
e-mail: aratyn@uic.edu\\
\and Johan van de Leur\\
\\
Mathematical Institute,\\
University of Utrecht,\\
P.O. Box 80010, 3508 TA Utrecht,\\
The Netherlands\\
e-mail: vdleur@math.uu.nl}

\newtheorem{lemma}{Lemma}[section]
\newtheorem{proposition}{Proposition}[section]
\newtheorem{theorem}{Theorem}[section]

\newtheorem{remark}{Remark}[section]

\def\lb{\lbrack}
\def\rb{\rbrack}
\def\({\left(}
\def\){\right)}
\def\tr{\mathop{\rm tr}}                  
\def\Tr{\mathop{\rm Tr}}
\def\th{\theta}
\def\a{\alpha}
\def\b{\beta}
\def\c{\chi}
\def\d{\delta}

\def\eps{\epsilon}

\def\g{\gamma}

\def\om{\omega}
\newcommand{\dder}[2]{\frac{d{#1}}{d{#2}}}
\newcommand{\pder}[2]{\frac{\partial{#1}}{\partial{#2}}}
\begin{document}

\maketitle
\begin{center}\it Dedicated to Pierre van Moerbeke on the occasion of his
$60^{th}$ birthday
\end{center}
\begin{abstract}
Equivalence  is established between a special class of Painleve VI equations
parametrized by a conformal dimension $\mu$, time dependent Euler top
equations,
isomonodromic deformations and three-dimensional Frobenius manifolds.
The isomodromic tau function and solutions of the Euler top equations are
 explicitly constructed in terms of  Wronskian
solutions of the 2-vector 1-constrained symplectic Kadomtsev-Petviashvili (CKP)
hierarchy by means of Grassmannian formulation.
These Wronskian solutions give rational solutions of the Painleve VI equation
for
$\mu=1,2,{\ldots} $.
\end{abstract}


\section[Painlev\'e VI  and Euler top equations]{Isomondromic deformation
problem, Painlev\'e VI  equation
and the Euler top equations}
Consider a Fuchsian system of linear differential equation with
rational coefficients:
\begin{equation}
\frac{\partial}{\partial z} \ X (a,z)= \ - \sum_{i=1}^3
\frac{A_i}{z-a_i} \ X (a,z) \ , \qquad
\frac{\partial}{\partial a_i}  \ X (a,z)=   \frac{A_i}{z-a_i} \ X (a,z)\,.
\label{eq:wone} 
\end{equation}
The three-dimensional Schlesinger equations
\begin{equation}
\frac{\partial}{\partial a_i} \ A_i =  \sum_{j=1,\ j\neq i}^3 \
\frac{ \left[ A_i,A_j \right] }{a_i -a_j} ,\quad
\frac{\partial}{\partial a_j} \ A_i =
\frac{ \left[ A_i, A_j \right]}{a_j-a_i} \ ,\qquad \ i \neq j \,.
\label{eq:paisi} 
\end{equation}
emerge as compatibility equations
of the system (\ref{eq:wone}) and
describe monodromy preserving deformations for the linear differential
equations in the complex plane.

Let us fix $a_1=0,a_2=1$ and $a_3=x$ and work with
$2 \times 2$ matrices $A_0,A_1,A_x$. The Schlesinger equations
reduce to:
\[
{\dot A}_0 = - \frac{1}{x} \left[ A_0,A_x \right],\quad
{\dot A}_1 = \frac{1}{1-x} \left[ A_1,A_x \right],\quad
{\dot A}_x = \frac{1}{x} \left[ A_0,A_x \right]- \frac{1}{1-x}\left[ A_1,A_x
\right],
\]
where ${\dot A}= d A/d x$.
The matrix $A_x$ can be eliminated by setting
$
A_x = - A_0 -A_1-A_\infty
$,
where $A_\infty = - \sum_{i=1}^3 A_i$ is an
integral of the Schlesinger equations (\ref{eq:paisi}).

We will now follow \cite{JM1}, \cite{JM2}, \cite{JM3}, 
see also \cite{mahoux} and describe a
connection to the Painlev\'e VI  equation.

Let $\pm \th_0/2, \pm \th_1/2, \pm \th_x/2, \pm \th_\infty /2$
be eigenvalues of $A_0,A_1,A_x$  and $A_\infty$ and so
\[
\tr (A_0^2) = \frac{1}{2} \theta_0^2, \;\;\;
\tr (A_1^2) = \frac{1}{2} \theta_1^2, \;\;\;\tr (A_x^2) = \frac{1}{2}
\theta_x^2, \;\;\;
\tr (A_\infty^2) = \frac{1}{2} \theta_\infty^2.
\]
We parametrize the traceless matrices $A_0,A_1$ as in \cite{JM1}, 
\cite{JM2}, \cite{JM3},
\cite{mahoux} :
\begin{equation}
A_i = \frac{1}{2} \begin{pmatrix} z_i & u_i (\theta_i-z_i)\\
(\theta_i+z_i)/{u_i}& -z_i
\end{pmatrix} , \quad i=0,1.
\label{eq:azomatrices}
\end{equation}
Following \cite{JM1}, \cite{JM2}, \cite{JM3}, \cite{mahoux} we replace $u_0$ and $u_1$ by
two new variables $k $ and $y$:
\begin{equation}
k = xu_0(z_0-\theta_0)-(1-x)u_1 (z_1-\theta_1), \quad
k y =xu_0 (z_0-\theta_0)
\label{eq:kydefs}
\end{equation}
as a result
the above isomonodromic deformation problem
leads to the Painlev\'e VI  equation  :
\begin{eqnarray}
{\ddot y} &=&\frac{1}{2}\left( \frac{1}{y}
+\frac{1}{y-1}+\frac{1}{y-x}\right)
{\dot {y}}^2- \left( \frac{1}{x} +\frac{1}{x-1}+\frac{1}{y-x}\right)
{\dot {y}}\nonumber \\
&  +& \frac{y(y-1)(y-x)}{x^2(x-1)^2}\left[
\a +\b\frac{x}{y^2}+\gamma\frac{x-1}{(y-1)^2}
+\delta \frac{x(x-1)}{(y-x)^2}\right] \, ,
\label{eq:pain6}
\end{eqnarray}
characterized by the parameters $(\a,\b,\g,\d)$
\[
\a = \frac{(1-\theta_\infty)^2}{2} , \;\;\;
\b = - \frac{\theta_0^2}{2} , \;\;\;
\g =  \frac{\theta_1^2}{2} , \;\;\;
\d =  \frac{1-\theta_x^2}{2}\,.
\]
We will at this point reduce a number parameters from four to two
by setting
$
\rho= \theta_0=\theta_1=\theta_x$ and
$\nu=\theta_\infty$.
These constants  $\rho$ and  $\nu $ parametrize $(\a,\b,\g,\d)$
as follows
\begin{equation}
\a = \frac{(1-\nu)^2}{2} , \;\;\;
\b = - \frac{\rho^2}{2} , \;\;\;
\g =  \frac{\rho^2}{2} , \;\;\;
\d =  \frac{1-\rho^2}{2}\, .
\label{eq:muparamtrization}
\end{equation}
In this formulation it convenient to define
\begin{eqnarray}
\om_1^2 &= &-\( \frac{\rho^2}{2} + \tr (A_1 A_x)\)=
-\frac{\rho^2}{4} -\frac{\nu^2}{4} - \frac{1}{2} \nu z_0
, \label{eq:oms1}\\
\om_2^2 &= &-\( \frac{\rho^2}{2} + \tr (A_0 A_1)\)=
-\frac{\rho^2}{4}+\frac{\nu^2}{4}+ \frac{1}{2} \nu  (z_0+z_1) ,
\label{eq:oms2}\\
\om_3^2 &= &-\( \frac{\rho^2}{2} + \tr (A_0 A_x)\)=
-\frac{\rho^2}{4}- \frac{\nu^2}{4}- \frac{1}{2} \nu  z_1 . \label{eq:oms3}
\end{eqnarray}
The functions $\om_i, i=1,2,3$ defined in (\ref{eq:oms1}-\ref{eq:oms3})
satisfy
\begin{equation}
\sum_{i=1}^3 \om_i^2 = -\frac{3 \rho^2}{4}- \frac{\nu^2}{4}
= - \mu^2\, ,
\label{eq:omsquare}
\end{equation}
which defines the scaling dimension $\mu$.

One can also prove like in \cite{H1}
that $\om_i$, $i=1,2,3$,  satisfy the time dependent Euler top equations:
\begin{equation}
\frac{d \om_1}{d x}= \frac{\om_2\om_3}{x}, \;\; \;\;\;\;
\frac{d \om_2}{d x}=  \frac{\om_1\om_3}{x(x-1)}, \;\;\;\;\;\;
\frac{d \om_3}{d x}= \frac{\om_1\om_2}{1-x} \, .
\label{eq:euta}
\end{equation}

Next, introduce
\begin{equation}
\zeta = x(1-y)z_0+(1-x)yz_1\, , .
\label{eq:zetadef}
\end{equation}
for which we have two equations \cite{JM1}, \cite{JM2}, \cite{JM3}, 
\cite{mahoux}:
\begin{eqnarray}
\zeta &=& -x (1-x) {\dot y}+(1-\theta_\infty) y (1-y),
\label{eq:zety}\\
2 \theta_\infty (z_0+z_1)  &=& 4 \om_2^2 +\rho^2-\nu^2
= \frac{ \rho^2 (y-x)^2 -\zeta^2}{x(1-x)y(1-y)}+\rho^2-\nu^2 .
\label{eq:xiy}
\end{eqnarray}
{}From which we can determine $\om_2^2 $ in terms of $y$ and ${\dot y}$
as
\[
\om_2^2 = \frac{\rho^2 (y-x)^2 -\zeta^2}{4 x(1-x)y(1-y)}\,.
\]
 {}From equations (\ref{eq:zetadef}), (\ref{eq:zety}) and (\ref{eq:xiy})
we can express $z_0$ or $z_1$ in terms of $y$ and ${\dot y}$.
This procedure yields:
\begin{eqnarray}
\om_1^2 &= &- \frac{\rho^2+\nu^2}{4} -
\frac{\rho^2-\nu^2}{4} \frac{(1-x)y}{y-x}
 +\frac{\nu^2}{2} \frac{y(y-1)}{y-x} \nonumber\\
&+&\nu \frac{A}{y-x} - \frac{A_{+}A_{-}}{x(y-1)(y-x)}
, \label{eq:oms1y}\\
\om_2^2 &= & \frac{A_{+}A_{-}}{x(1-x)y(y-1)}
, \label{eq:oms2y}\\
\om_3^2 &= &- \frac{\rho^2+\nu^2}{4} -
\frac{\rho^2-\nu^2}{4} \frac{x(1-y)}{y-x}
-\frac{\nu^2}{2} \frac{y(y-1)}{y-x}\nonumber\\
&-&\nu \frac{A}{y-x} - \frac{ A_{+}A_{-}}{(1-x)y(y-x)}
, \label{eq:oms3y}
\end{eqnarray}
where
\begin{eqnarray}
A &=& \frac{1}{2} \left[ {\dot y} x (x-1) - y (y-1) \right] ,\label{eq:adef}\\
A_{\pm} &=& \frac{1}{2}  {\dot y} x (x-1)- \frac{1}{2} (1-\theta_\infty) y
(y-1) \pm
\frac{1}{2} \rho (y-x) \nonumber \\
 &=& A+ \frac{1}{2} \nu y (y-1)\pm
\frac{1}{2} \rho  (y-x)
\, . \label{eq:apmdef}
\end{eqnarray}

There are two natural ways to further reduce  the system
to a one parameter system characterized by a conformal
scaling dimension $\mu$ only.

{\bf 1)} Set $\rho^2=\nu^2$. Thus, from (\ref{eq:omsquare})
$\rho^2=\nu^2=\mu^2$ with (c.f. \cite{H1,H2,H3})
\begin{equation}
\alpha= \frac{(1\mp \mu)^2}{2},\;\;\; \beta = - \frac{\mu^2}{2},\;\;\;
\gamma= \frac{\mu^2}{2},\;\;\; \delta = \frac{1-\mu^2}{2},
\label{eq:reduca}
\end{equation}
using that $\nu = \pm \mu$. For instance, for
$\nu =1/2$ we get
$(\a,\b,\g,\d) = ( 1/8,-1/8,1/8,3/8)$, while
for $\nu=-1/2$ we get
$(\a,\b,\g,\d) = ( 9/8,-1/8,1/8,3/8)$.
In this case $\om_i, i=1,2,3$ are defined through
(\ref{eq:oms1})-(\ref{eq:oms3}):
\begin{equation}
\label{eq:omsn}
\om_1^2 =
-\frac{\mu^2}{2} - \frac{1}{2} \nu  z_0,\quad
\om_2^2 =
 \frac{1}{2} \nu  (z_0+z_1) ,\quad
\om_3^2 =
-\frac{\mu^2}{2} - \frac{1}{2} \nu  z_1,
\end{equation}
which now yields :
\begin{eqnarray}
\om_1^2 &= &- \frac{\mu^2}{2} \(1  + \frac{y(1-y)}{y-x}\)
+\nu  \frac{A}{y-x} - \frac{1}{x(y-1)(y-x)} A_{+}A_{-}
, \label{eq:ooms1y}\\
\om_2^2 &= & \frac{1}{x(1-x)y(y-1)} A_{+}A_{-}
, \label{eq:ooms2y}\\
\om_3^2 &= &- \frac{\mu^2}{2}\(1  - \frac{y(1-y)}{y-x}\)
-\nu \frac{A}{y-x} - \frac{1}{(1-x)y(y-x)} A_{+}A_{-}
, \label{eq:ooms3y}
\end{eqnarray}
where $A$ is as in (\ref{eq:adef}) and
\begin{equation}
A_{\pm} = \frac{1}{2}  {\dot y} x (x-1)- \frac{1}{2} (1-\nu ) y (y-1) \pm
\frac{1}{2} \mu  (y-x)
 = A+ \frac{1}{2} \nu y (y-1)\pm\frac{1}{2} \mu (y-x)
, \label{eq:apmndef}
\end{equation}
with $\nu = \pm \mu$.

For $\nu=1/2$ (and $\mu^2=1/4$) expressions (\ref{eq:oms1y}-\ref{eq:oms3y})
agree with results of \cite{UNESP}.

{}From equations (\ref{eq:zetadef}) and (\ref{eq:xiy}) we find
for $\rho^2=\nu^2=\mu^2$:
\begin{equation}
\mu^2 \om_2^2 x(1-x)y(1-y) - \mu^4 (y-x)^2/4 +
\left[ x (1-y) (\om_1^2+\mu^2/2)+(1-x)y
(\om_3^2+\mu^2/2)\right]^2=0,
\label{eq:ddoms2your}
\end{equation}
which yields a solution of the Painlev\'e VI equation
of the form :
\begin{equation}
y(x) = x \frac{\pm (x-1) \mu \omega_1 \omega_2 \omega_3 + x \omega_1^2
\omega_2^2+\omega_1^2 \omega_3^2}{(x-1)^2 \omega_2^2 \omega_3^2 +x^2
\omega_1^2 \omega_2^2+\omega_1^2 \omega_3^2 }.
\label{eq:ypour}
\end{equation}

{\bf 2)} In the second case we set $\rho =0$ and therefore from
(\ref{eq:omsquare})
$\nu^2 = (2\mu)^2$ with (c.f. \cite{Du2,DubMaz,mazzocco})
the result that
\begin{equation}
\alpha= \frac{(1 \pm 2\mu)^2}{2},\;\;\; \beta = 0,\;\;\;
\gamma= 0,\;\;\; \delta = \frac{1}{2}
\label{eq:reducb}
\end{equation}
and (see (\ref{eq:oms1}-\ref{eq:oms3}))
\begin{equation}
\om_1^2 =
- \mu^2 - \frac{1}{2} \nu z_0
,\quad
\om_2^2 =
+\mu^2+ \frac{1}{2} \nu  (z_0+z_1) , \quad
\om_3^2 =
- \mu^2- \frac{1}{2} \nu  z_1 ,
\end{equation}
which now yields
\begin{eqnarray}
\om_1^2 &= &-  \frac{(y-1)(y-x)}{x} \left[ \frac{A}{(y-1)(y-x)}
+ \frac{\nu}{2} \right]^2
, \label{eq:doms1y}\\
\om_2^2 &= & \frac{y(y-1)}{x(1-x)} \left[ \frac{A}{y(y-1)}
+ \frac{\nu}{2} \right]^2
, \label{eq:doms2y}\\
\om_3^2 &= &- \frac{y(y-x)}{(1-x)} \left[ \frac{A}{y(y-x)}
+ \frac{\nu}{2} \right]^2 .
 \label{eq:doms3y}
\end{eqnarray}
{}From (\ref{eq:xiy}) we find for $\rho=0$
\[
 \om_2^2
= -\frac{ \zeta^2}{4x(1-x)y(1-y)},
\]
which together with definition (\ref{eq:zetadef}) of $\zeta$
and (\ref{eq:doms1y}) and (\ref{eq:doms3y}) yields equation
\begin{equation}
4 \mu^2 \om_2^2 x(1-x)y(1-y) + \left[ x (1-y) (\om_1^2+\mu^2)+(1-x)y
(\om_3^2+\mu^2)\right]^2=0.
\label{eq:ddoms2y}
\end{equation}

As a general solution of (\ref{eq:ddoms2y}) one obtains
expressions
\begin{equation}
y(x) = - x \frac{x(\om_1\om_2 \mp \mu\om_3)^2+(\om_1\om_3 \pm \mu\om_2)^2}{
\(\om_3^2+\mu^2+x (\om_2^2+\mu^2)\)^2+4 x \mu^2\om_1^2} .
\label{eq:ydubr}
\end{equation}

As an example we consider the case of
$\mathbf{ \mu}= \pm 1$
with
\begin{equation}
\om_1 = {\displaystyle \frac {\sqrt{ - b}\,(1 - x)}{b - x}}
, \quad
\om_2 =  - {\displaystyle \frac {\sqrt{ - b\,(b - 1)}}{b- x}}
, \quad
\om_3 = {\displaystyle \frac {\sqrt{b - 1}\,x}{b - x}} ,
\label{eq:omeone}
\end{equation}
which satisfy the Euler top equations (\ref{eq:euta}) and
$
\sum_{i=1}^3 \om_i^2 = -1 $, hence
$\mu^2 =1$.
As one of two solutions to equation (\ref{eq:ddoms2y})
we obtain
\begin{equation}
y (x) =  - {\displaystyle \frac {(b - 1)\,x}{ - b + x}} ,
\label{eq:ypain6mu1}
\end{equation}
which satisfies the Painlev\'e VI equation
(\ref{eq:pain6}) with
\[
(\a,\b,\g,\d)= \( (1-2 \mu)^2/2,0,0,1/2\)= \(1/2,0,0,1/2\),
\]
corresponding to $\mu=1$.
Note, that introducing $ a= (b-1)/b, a\ne 0$ we can rewrite
(\ref{eq:ypain6mu1}) as
\[
y (x) =   {\displaystyle \frac {a\,x}{1 - (1-a) x}} ,
\]
which appeared in \cite{mazzocco}
as a one parameter family of rational solutions to
Painlev\'e VI equation with ${\mu=1}$.

As a second solution to equation (\ref{eq:ddoms2y}) we obtain
for (\ref{eq:omeone})
\[
y(x) :=  - {\displaystyle \frac {x\,(b - 1)\,( - b + x^{2}
)^{2}}{( - b + x)\,(x^{4} - 4\,b\,x^{3} + 6\,b\,x^{2} - 4\,b\,x
 + b^{2})}} ,
\]
which satisfies the Painlev\'e VI equation
(\ref{eq:pain6}) with
\[
(\a,\b,\g,\d)= \( (1-2 \mu)^2/2,0,0,1/2\)= \(9/2,0,0,1/2\).
\]
corresponding to $\mu=-1$.

There is only one solution of equation (\ref{eq:ddoms2your}):
\[
y(x) = {\displaystyle \frac {x^{2} - b}{2\,( - b + x)}} ,
\]
which yields a solution of the Painlev\'e VI equation
(\ref{eq:pain6}) with
\[
(\a,\b,\g,\d)= \( (1\pm  \mu)^2/2,- \mu^2/2,\mu^2/2,(1-\mu^2)/2\)
= \(2,-1/2,1/2,0\).
\]

\section{The Darboux-Egoroff equations}
\label{Sec1}
The connection between the Painlev\'e VI equation and three-dimensional
Frobenius manifolds
is established through
the Darboux-Egoroff equations for the rotation coefficients $\beta_{ij} =
\beta_{ji}$:
\begin{equation}
    \frac{\partial }{ \partial u_k} \beta_{ij} = \beta_{ik} \beta_{kj},
\;\; \;\mbox{distinct}\;\; i,j,k,
\label{betas-comp}
\end{equation}
\begin{equation}
\sum_{k=1}^{n} \frac{\partial }{ \partial u_k} \beta_{ij}     =0,
\;\; i \ne j \, .
\label{ionb}
\end{equation}
In addition to these equations one also assumes the conformal condition:
\begin{equation}
\sum_{k=1}^nu_k\frac{\partial }{ \partial u_k} \beta_{ij} = -\beta_{ij}\, .
\label{betas-deg}
\end{equation}
 The Darboux-Egoroff equations (\ref{betas-comp})-(\ref{ionb})
appear as compatibility equations of a linear system :
\begin{eqnarray}
\frac{\partial \Phi_{ij} (u,z) }{\partial u_k} &=& \beta_{ik} (u) \Phi_{kj}
(u,z)
\;\;\; i \ne k,\label{delinsa} \\
\sum_{k=1}^n \frac{\partial \Phi_{ij} (u,z) }{\partial u_k} &=&
z \Phi_{ij}  (u,z).
\label{delinsb}
\end{eqnarray}
Define the $n \times n$ matrices $\Phi =(\Phi_{ij})_{1\le i,j\le n}$,
$B =(\beta_{ij})_{1\le i,j\le n}$ and
$ V_i = \left\lbrack B \, ,\, E_{ii} \right\rbrack$, where
$(E_{ij})_{k\ell}=\delta_{ik}\delta_{j\ell}$.
Then the linear system (\ref{delinsa})-(\ref{delinsb})
acquires the following form :
\begin{eqnarray}
\frac{\partial \Phi (u,z) }{\partial u_i} &=&
\left( z E_{ii} + V_i (u) \right)\Phi (u,z), \quad
i=1,{\ldots} ,n \;, \label{delinsma} \\
\sum_{k=1}^n \frac{\partial \Phi (u,z) }{\partial u_k} &=&
z \Phi  (u,z) \, .
\label{delinsmb}
\end{eqnarray}

The  conformal case $n=3$ is very special. In that case
\begin{equation}
V=[B,U]=[
\begin{pmatrix}
0&\beta_{12}&\beta_{13}\\
\beta_{21}&0&\beta_{23}\\
\beta_{31}&\beta_{32}&0
\end{pmatrix},\begin{pmatrix}
u_1&0&0\\
0&u_2&0\\
0&0&u_3
\end{pmatrix}]=\begin{pmatrix}
0&\omega_3&-\omega_2\\
-\omega_3&0&\omega_1\\
\omega_2&-\omega_1&0
\end{pmatrix}
\label{eq:vome}
\end{equation}
satisfies
\begin{equation}
\pder{V}{u_j} = \lbrack V_j \, , \, V \rbrack .
\label{eq:pajv}
\end{equation}

Note, that $\Tr (V^2)$ is an integration constant
of equations (\ref{eq:pajv}), as it follows easily that
$\partial \Tr (V^2)/\partial u_j =0$ for all $j$.

For three-dimensional Frobenius manifolds, these equations
exhibit isomonodromic dependence  on canonical coordinates
$u$ and reduce to the class of the Painlev\'e VI equation
(\ref{eq:pain6}) with $(\a,\b\,g,\d)$ parameters as in
(\ref{eq:reduca}) or (\ref{eq:reducb}).

For vectorfields $I = \sum_{j=1}^3 \partial / \partial u_j$
and $E = \sum_{j=1}^3 u_j \partial / \partial u_j$ it follows
from (\ref{eq:pajv}) that $I (V)=0, E(V)=0$ and
accordingly $V$ is a function
of one variable $x$ such that $I (x)=0, E(x)=0$.
We choose
\begin{equation}
x= \frac{u_2-u_1}{u_3-u_1}\,.
\label{eq:tdef}
\end{equation}

Note that $\tr (V)=0$ and $\det (V) =0$ and $V$ has
eigenvalues $\mu,0,-\mu$ where $\mu$ defines
the integration
constant $\Tr (V^2)$ of \ref{eq:pajv} through :
\[
\Tr (V^2) = - 2 \( \om_1^2+\om_2^2+\om_3^2\) = 2 \mu^2\, .
\]
Then $\om_i, i=1,2,3$
satisfy the Euler top equations (\ref{eq:euta}) as a result of
(\ref{eq:pajv}).

Note that  $V(x)$, $V(u_1,u_2,u_3)$, i.e. $V$ as function of $x$,
respectively the  $u_i$'s, are connected as follows
\[
V(x)=V(0,x,1),\qquad V(u_1,u_2,u_3)=V\left( \frac{u_2-u_1}{u_3-u_1}\right).
\]
Since
\[
\begin{aligned}
\omega_1(u_1,u_2,u_3)=(u_3-u_2)\beta_{32}(u_1,u_2,u_3),\\[2mm]
\omega_2(u_1,u_2,u_3)=(u_1-u_3)\beta_{13}(u_1,u_2,u_3),\\[2mm]
\omega_3(u_1,u_2,u_3)=(u_2-u_1)\beta_{12}(u_1,u_2,u_3).
\end{aligned}
\]
We find that
\[
\omega_1(x)=(1-x)\beta_{23}(0,x,1),\qquad
\omega_2(x)=-\beta_{13}(0,x,1),\qquad
\omega_3(x)=x\beta_{12}(0,x,1).
\]
In other words, it suffices to know the rotation coefficients
$\beta_{ij}(0,x,1)$.

\section{The tau-function}
We define the $\tau$-function by
equation:
\begin{equation}
\pder{\log \tau}{u_j} =  \frac{1}{2} \Tr \(V_j V\) \; = \;
\sum_{i=1}^3 \b_{ij}^2 (u_i-u_j)
 =  \sum_{i,k=1}^3 \eps_{ijk}^2 \frac{\om_k^2}{(u_i-u_j)}
\,,
\label{eq:pajltbu}
\end{equation}
in which we used $\b_{ij} = \eps_{ijk} \om_k /(u_j-u_i)$.
This tau-function is related as
\[
\tau_I=\frac{1}{\sqrt \tau}
\]
to Dubrovin's isomonodromy tau-function
$\tau_I$ \cite{dubrov}.

The identity $I (\log \tau) =0$, shows that $\tau$ is a function of two
variables, which again can be identified with $t$ and $h$
such that
\begin{equation}
h=u_2-u_1 \, .
\label{eq:defh}
\end{equation}
It follows from (\ref{eq:pajltbu}) that
\[
E \( \log \tau (u) \) =\frac{1}{2} \Tr \(V^2\)= \mu^2\,.
\]
Making use of technical identities :
\[
\pder{x}{u_1}=\frac{1}{h} (x-1)x, \quad
\pder{x}{u_2}=\frac{1}{h} x, \quad
\pder{x}{u_3}=- \frac{1}{h} x^2 ,
\]
one easily derives
\[
\pder{}{u_1} = \frac{x(x-1)}{h} \pder{}{x}-\pder{}{h},\;\;\; \,
\pder{}{u_2}= \frac{x}{h}\pder{}{x}+\pder{}{h},\;\;\; \,
\pder{}{u_3}=-\frac{x^2}{h}\pder{}{x}\, ,
\]
{}from which
\[
E= h \pder{}{h}\,
\]
follows.
Since $ E(\log \tau)= h \partial \log \tau / \partial h = \mu^2$ we see that
$\log \tau (x,h)$ decomposes  as
\begin{equation}
\log \tau (x,h) = \mu^2 \log h +\log \tau_0 (x)
\label{eq:tau-decom}
\end{equation}
where $\tau_0$ is a function of $x$ only.

It follows from equations \ref{eq:pajv}
and \ref{eq:pajltbu} that
\[
\frac{\partial^2 \log\tau }{\partial u_i \partial u_j}=
- \beta^2_{ij}  , \qquad i \ne j\, .
\]
which translates to a following parametrization
of $\omega_i$'s in terms of a single isomonodromic tau function :
\begin{equation}
\begin{split}
\omega_2^2 &= x\,(x - 1)\,({\frac {d^{2}}{dx^{2}}}\,\mathrm{ln}(
\tau_0 )(x)) + (2\,x - 1)\,({\frac {d}{dx}}\,\mathrm{ln}(\tau_0 )(x))
=\dder{}{x} \left\lb x (x-1) {\frac {d}{dx}}\,\mathrm{ln}(\tau_0 )(x)
\right\rb ,\\
\omega_3^2 &=  - x^{2}\,(x - 1)\,({\frac {d^{2}}{dx^{2}}}\,
\mathrm{ln}(\tau_0 )(x)) - x^{2}\,({\frac {d}{dx}}\,\mathrm{ln}(
\tau_0 )(x)) - \mu ^{2}
= -x^2 \dder{}{x} \left\lb  (x-1) {\frac {d}{dx}}\,\mathrm{ln}(\tau_0 )(x)
\right\rb - \mu ^{2},
\\
\omega_1^2 &= x\,(x - 1)^{2}\,({\frac {d^{2}}{dx^{2}}}\,\mathrm{ln
}(\tau_0 )(x)) + (x - 1)^{2}\,({\frac {d}{dx}}\,\mathrm{ln}(\tau_0 )(
x))=
(x-1)^2 \dder{}{x} \left\lb  x {\frac {d}{dx}}\,\mathrm{ln}(\tau_0 )(x)
\right\rb .
\end{split}
\label{eq:omegas}
\end{equation}
One verifies that indeed $
\omega_1^2 +\omega_2^2 +\omega_3^2= - \mu^2$.
Moreover, \[
\dder{\ln \tau_0}{x} = \frac{\omega_1^2}{x(1-x)} +\frac{\omega_2^2}{x}.
\]

\section{The CKP hierarchy}
\label{Sec2}

The symplectic Kadomtsev-Petviashvili or CKP hierarchy \cite{Date81} can be
obtained as a reduction of the KP
hierarchy,
\begin{equation}
\label{1}
\frac{\partial }{ \partial t_n} { L}  = \lbrack ({
L}^n)_{+}\, , \,  { L}  \rbrack ,\qquad\mbox{for }
{ L} =L(t,\partial)= {\partial } + \ell^{(-1)}(t){\partial }^{-1}+
\ell^{(-2)}(t){\partial }^{-2}+\cdots,
\end{equation}
where $x=t_1$ and $\partial=\frac{\partial}{\partial x}$, by assuming the extra
condition
\begin{equation}
\label{2}
L^*=-L.
\end{equation}
By taking the adjoint, i.e.,$\ ^*$  of (\ref{1}),
one sees that $\frac{\partial L}{\partial t_n}=0$ for $n$ even. Date, Jimbo,
Kashiwara and Miwa \cite{Date81}, \cite{JM} construct such $L$'s from certain
special KP wave functions  $\psi(t,z)=P(t,z)e^{\sum_it_iz^i}$
(recall $L(t,\partial)=P(t,\partial)\partial  P(t,\partial)^{-1}$), where one
then puts all even times $t_n$ equal to 0.
Recall that a KP wave function satisfies
\begin{equation}
\label{2a}
L\psi(t,z)=z\psi(t,z),\qquad
\frac{\partial \psi(t,z)}{ \partial t_n}= (
L^n)_{+} \psi(t,z),
\end{equation}
and
\begin{equation}
\label{2b}
Res_z\, \psi(t,z)\psi^*(s,z)=0.
\end{equation}
The special wave functions which lead to an $L$ that has condition (\ref{2})
satisfy
\begin{equation}
\label{3}
\psi^*(t,z)=\psi(\tilde t,-z),
\qquad \mbox{where }\quad \tilde t_i=(-)^{i+1}t_i.
\end{equation}
We call such a $\psi$ a  CKP wave function. Note that this implies that
$L(t,\partial)^*=-L(\tilde t,\partial)$
and that
\[
Res_z \,\psi(t,z)\psi(\tilde s,-z)=0.
\]
One can put all even times equal to 0, but we will not do that here.

The CKP wave functions correspond to very special points in the
Sato Grassmannian, which consists of all linear spaces
\[
W\subset
H_+\oplus H_-={\mathbb C}[z]\oplus z^{-1}{\mathbb C}[[z^{-1}]],
\]
such that the projection on $H_+$ has finite index. Namely,
$W$ corresponds to a CKP wave function if the index is 0 and for any
$f(z),g(z)\in W$ one has $Res_z\, f(z)g(-z)=0$. The corresponding
CKP tau functions satisfy $\tau(\tilde t)=\tau(t)$.

We will now generalize this to the multi-component case and show that a
CKP reduction of the multi-component KP hierarchy gives the Darboux-Egoroff
system. The $n$ component
KP hierarchy  \cite{ANP99}, \cite{KvdL} consists of the equations in
$t_j^{(i)}$, $1\le i\le n$,
$j=1,2,\ldots$
\begin{equation}
\label{4}
\frac{\partial }{ \partial t_j^{(i)}} { L}  = \lbrack ({
L}^jC_i)_{+}\, , \,  { L}  \rbrack ,\qquad
\frac{\partial }{ \partial t_j^{(i)}} { C_k}  = \lbrack ({
L}^jC_i)_{+}\, , \,  { C_k}  \rbrack ,
\end{equation}
for the commuting $n\times n$-matrix pseudo-differential operators,
$L$, $C_i$, $i=1,2,\ldots n$, with $\sum_i C_i=I$ of the form
\begin{equation}
{ L} = {\partial } + L^{(-1)}{\partial }^{-1}+
L^{(-2)}{\partial }^{-2}+\cdots,\qquad C_i=E_{ii}+C_i^{(-1)}{\partial }^{-1}+
C_i^{(-2)}{\partial }^{-2}+\cdots,
\end{equation}
$1\le i\le n$, where
$x=t_1^{(1)}+t_1^{(2)}+\cdots+t_1^{(n)}$.
The corresponding wave function has the form
\[
\Psi(t,z)=P(t,z)\exp \left(\sum_{i=1}^n\sum_{j=1}^\infty
t_j^{(i)}z^j E_{ii}\right),\quad\mbox{where }
P(t,z)=I+P^{(-1)}(t)z^{-1}+\cdots,
\]
and satisfies
\begin{equation}
\label{5}
L\Psi(t,z)=z\Psi(t,z),\quad C_i\Psi(t,z)=\Psi(t,z)E_{ii},\quad
\frac{\partial \Psi(t,z)}{ \partial t_j^{(i)}}= (
L^jC_i)_{+} \Psi(t,z)
\end{equation}
and
\[
Res_z\, \Psi(t,z)\Psi^*(s,z)^T=0.
\]
{}From this we deduce that $L=P(t,\partial  )\partial P(t,\partial  )^{-1}$
and
$C_i=P(t,\partial  )E_{ii}P(t,\partial  )^{-1}$.
Using this, the simplest equations in (\ref{5}) are
\begin{equation}
\label{V0}
\frac{\partial \Psi(t,z)}{ \partial t_1^{(i)}}= (zE_{ii}+V_i(t))
\Psi(t,z),
\end{equation}
where $V_i(t)=[B(t), E_{ii}]$ and $B(t)=P^{(-1)}(t)$.
In terms of the matrix coefficients $\beta_{ij}$ of $B$ we obtain
(\ref{betas-comp})  for  $u_i=t^{(i)}_1$.

The Sato Grassmannian becomes vector valued, i.e.,
\[
H_+\oplus H_-=({\mathbb C}[z])^n\oplus z^{-1}({\mathbb C}[[z^{-1}]])^n.
\]
The same restriction as in the 1-component case (\ref{3}), viz.,
\[
\Psi(t,z)=\Psi^*(\tilde t,-z),
\qquad \mbox{where }\quad \tilde t_n^{(i)}=(-)^{n+1}t_n^{(i)}.
\]
leads to
$L^*(\tilde t)=-L(t)$, $C_i^*(\tilde t)=C_i(t)$ and
\begin{equation}
\label{A2}
Res_z\, \Psi(t,z)\Psi(\tilde s,-z)^T=0,
\end{equation}
which we call the multi-component CKP hierarchy.
But more importantly, it  also gives the restriction
\begin{equation}
\label{DB2}
\beta_{ij}(t)=\beta_{ji}(\tilde t).
\end{equation}
Such CKP wave functions correspond to points $W$ in the Grassmannian for which
\[
Res_z\, f(z)^Tg(-z)=Res_z\, \sum_{i=1}^n f_i(z)g_i(-z)=0
\]
for any $f(z)=(f_1(z),f_2(z),\ldots, f_n(z))^T,\
g(z)=(g_1(z),g_2(z),\ldots, g_n(z))^T\in W$.

If we finally assume that $L=\partial $, then $\Psi$, $W$
also satisfy
\begin{equation}
\label{A1}
\frac{\partial \Psi(t,z)}{\partial x}=\sum_{i=1}^n\frac{\partial
\Psi(t,z)}{\partial t_1^{(i)}}=z\Psi(t,z),\qquad zW\subset W
\end{equation}
and  thus $\beta_{ij}$ satisfies (\ref{ionb}) for $u_i=t_1^{(i)}$. Now
differentiating (\ref{A2}) $n$ times to $x$  for $n=0,1,2,\ldots$
and applying (\ref{A1}) leads to
\[
\Psi(t,z)\Psi(\tilde t,-z)^T=I.
\]
These special  points in the Grassmannian can all be constructed as follows
\cite{vdL1}.
Let $G(z)$ be an
element in  $GL_n({\mathbb C}[z,z^{-1}])$ that satisfies
\begin{equation}
\label{twist2}
G(z)G(-z)^T=I,
\end{equation}
then $W=G(z)H_+$. Clearly, any two $f(z),
g(z)\in W$ can be written as $f(z)=G(z)a(z),\ g(z)=G(z)b(z)$ with
$a(z), b(z)\in
H_+$, then $
zf(z)=zG(z)a(z)=G(z)za(z)\in W$, since $za(z)\in H_+$. Moreover,
\[
Res_z\, f(z)^Tg(-z)=Res_z\, a(z)^TG(z)^TG(-z)b(-z)=Res_z\, a(z)^Tb(-z)=0.
\]
We now take very special elements in this twisted loop
group, i.e., elements that correspond to certain
points of the Grassmannian that have a basis of
homogeneous elements in $z$.
Choose integers $\mu_1\le \mu_2\le \cdots\le \mu_n$
such that
$\mu_{n+1-j}=-\mu_j$.
Then take $G(z)$ of the form
\[
G(z)=N(z)S^{-1}=Nz^{-\mu}S^{-1},\qquad\mbox{where}\quad
\mu=\mbox{diag}({\mu_1},{\mu_2}, \ldots,{\mu_n})
\]
and $N=(n_{ij})_{1\le i,j \le n}$ a constant matrix that satisfies
\begin{equation}
\label{N}
N^TN=\sum_{j=1}^n (-1)^{\mu_j}E_{j,n+1-j}
\end{equation}
and
\[
S=\delta_{n,2m+1}E_{m+1,m+1}+\sum_{j=1}^m \frac{1}{\sqrt 2}
\left( E_{jj}+iE_{n+1-j,j}+
E_{j,n+1-j}-iE_{n+1-j,n+1-j}\right),
\]
for $n=2m$ or $n=2m+1$.
Then \cite{AL1}
\[
\sum_{i=1}^n\sum_{j=1}^\infty jt_j^{(i)}\frac{\partial \Psi(t,z)}{\partial
t_j^{(i)}}=z\frac{\partial \Psi(t,z)}{\partial
z},
\]
from which one deduces that
\begin{equation}
\label{conf-b}
\sum_{i=1}^n\sum_{j=1}^\infty jt_j^{(i)}\frac{\partial \beta_{ij}}{\partial
t_j^{(i)}}=-\beta_{ij}.
\end{equation}

We next put $t_j^{(i)}=0$ for all $i$ and all $j>1$
and $u_i=t^{(i)}_1$, then we obtain the situation of Section \ref{Sec1}.

\section{The case n=3}
\label{Sec3}

We will now give an example of the previous construction, viz., the case that
$n=3$
 and
$-\mu_1=\mu_3=\mu\in\mathbb{N}$ and $\mu_2=0$. Hence, the point of the
Grassmannian is given
by
\[
N(z)H_+= N\begin{pmatrix} z^{-\mu}&0&0\\ 0&1&0\\ 0&0&z^{\mu}
\end{pmatrix}H_+.
\]
More precise, let $n_i=(n_{1i},n_{2i},n_{3i})^T$ and $e_1=(1,0,0)^T$,
$e_2=(0,1,0)^T$ and $e_3=(0,0,1)^T$, then this point of the
Grassmannian has as basis
\[
\begin{aligned}
n_1z^{-\mu},\ n_1z^{1-\mu},\ \ldots,\ n_1z^{-1}, n_1,\ n_2,\ n_1z,\ n_2z,\
\ldots,\ ,n_1z^{\mu-1},\ n_2z^{\mu-1},\\[2mm] e_1z^\mu,\ e_2z^\mu,\ e_3z^\mu,\
e_1z^{\mu+1},\ e_2z^{\mu+1},\ \cdots &.
\end{aligned}
\]
Using this one can calculate in a similar way as in \cite{LM} (using the
boson-fermion correspondence or vertex operator constructions) the wave
function:
\[
\begin{aligned}
\Psi(t,z)=&P(t,z)\exp \left(\sum_{i=1}^n\sum_{j=1}^\infty
t_j^{(i)}z^j E_{ii}\right),\quad\mbox{where }\\
P_{jj}(t,z)=&\frac{\hat\tau(t^{(k)}_\ell-\delta_{kj}(\ell
z^\ell)^{-1})}{\hat\tau(t)},\quad
P_{ij}(t,z)=z^{-1}\frac{\hat\tau_{ij}(t^{(k)}_\ell-\delta_{kj}(\ell
z^\ell)^{-1})}{\hat\tau(t)}\quad \mbox{for }i\ne j
\end{aligned}
\]
and where
\[
\hat\tau(t)=\det\sum_{k=1}^3\sum_{i=0}^{\mu-1}\left(
\sum_{j=1}^{2\mu}n_{k1}S_{\mu+i-j+1}(t^{(k)}) E_{3i+k,j}+
\sum_{j=1}^{\mu}n_{k2}S_{i-j+1}(t^{(k)}) E_{3i+k,2\mu+j}\right).
\]
The functions  $S_i(x)$ are the elementary Schur polynomials, defined by:
\[
\sum_{j\in\mathbb{Z}}S_j(x)z^j=e^{\sum_{k=1}^\infty x_kz^k}.
\]
The  tau function
$\hat\tau_{ij}(t)$ is up to a multiplicative factor -1 equal to the above
determinant
where we replace the $j$-th row by
\[
\begin{pmatrix}
n_{i1}S_{\mu-1}(t^{(i)})&\cdots&n_{i1}S_1(t^{(i)})&n_{i1}&0&\cdots&0&0
\end{pmatrix}.
\]
Then
\begin{equation}
\label{bt}
\beta_{ij}(t)=\frac{\hat\tau_{ij}(t)}{\hat\tau(t)}.
\end{equation}

As we have seen in section \ref{Sec1} it suffices to calculate $\beta_{ij}(t)$
only for $t_1^{(2)}=x$, $t_1^{(3)}=1$   However we will not do that yet,
we will take $t_1^{(j)}=s_j$, for $j=1,2,3,\ldots$, $t_1^{(3)}=1$ and all other
$t_i^{(j)}=0$   and write $\beta_{ij}(s)$ for the resulting $\beta_{ij}$. In
fact we will make this substitution
in $\hat\tau(t)$ and $\hat\tau_{ij}(t)$. This might lead to
$\hat\tau(s)=\hat\tau_{ij}(s)=0$ in such a way that
$\beta_{ij}(s)=\frac{\hat\tau_{ij}(s)}{\hat\tau(s)}\ne 0$. However, as we shall
see later, this will not happen.

Since,  we can multiply the columns of the matrices of $\hat\tau_{ij}(s)$ and
$\hat\tau(s)$
by a constant we can change the
vectors $n_1=(n_{11},n_{21},n_{31})^T$ and  $n_2=(n_{12},n_{22},n_{32})^T$.
This will multiply $\hat\tau(s)$ by a scalar, but
also $\hat\tau_{ij}(s)$ by the same scalar, hence $\beta_{ij}(s)$ remains the
same. In a similar way $\beta_{ij}(t)$ does
not change if we permute the rows of $\hat\tau(s)$ and $\hat\tau_{ij}(s)$ in
the same way.
We thus choose
\[
n_1=(\alpha,1,a)^T,\qquad n_2=(-a,0,\alpha)^T,\qquad\mbox{with } \alpha, a\ne
0\ \mbox{and }\alpha^2+a^2=-1.
\]
Then our new $\hat\tau(s)$ becomes:
\[
\begin{aligned}
\hat\tau(s)=\det \sum_{i=1}^\mu
&\left(  \alpha E_{i,\mu+i}-aE_{i,2\mu+i}+\sum_{j=1}^{2\mu}
S_{\mu+i-j}(s)E_{\mu+i,j}+\frac{a}{(\mu+i-j)!}E_{2\mu+i,j}\right.\\[2mm]
&\qquad
\qquad\qquad\qquad\qquad\left.+\sum_{j=1}^\mu\frac{\alpha}{(i-j)!}
E_{2\mu+i,2\mu+j}\right)
,\end{aligned}
\]
where we assume that $k!=\infty$ for $k<0$.
And
$\hat\tau_{12}(s)$,  $\hat\tau_{13}(s)$ and $\hat\tau_{32}(s)$ is $-1$ times
the same
determinant, but now with the $\mu+1$-th, $2\mu+1$-th, $\mu+1$-th row,
respectively, replaced by
\[
\begin{aligned}
\left(
\begin{array}{cccc|ccc|ccc}
0&\cdots&0&\alpha&0&\cdots&0&0&\cdots&0
\end{array}
\right),\qquad
\left(
\begin{array}{cccc|ccc|ccc}
0&\cdots&0&{\alpha}&0&\cdots&0&0&\cdots&0
\end{array}
\right),\\[2mm]
\left(
\begin{array}{cccc|ccc|ccc}
\frac{a}{(\mu-1)!}&\frac{a}{(\mu-2)!}&\cdots&\frac{a}{0!}&0&\cdots&0&0&\cdots&0
\end{array}
\right),\quad\mbox{respectively.}
\end{aligned}
\]
Next subtract a multiple of the $2\mu+j$-th column from the $\mu+j$-th column,
then one sees that
\begin{equation}
\label{hattau}
\hat\tau(s)=\det \sum_{i=1}^\mu \left(E_{\mu+i,\mu+i}
+\sum_{j=1}^\mu S_{\mu+i-j}(s)E_{ij}-a^2S_{i-j}(s)E_{i,\mu+j}
+\sum_{j=1}^{2\mu}\frac{1}{(\mu+i-j)!}E_{\mu+i,j}\right)
,
\end{equation}
and  $\hat\tau_{12}(s)$, $\hat\tau_{13}(s)$, $\hat\tau_{32}(s)$, respectively
 is  the same determinant with the $1$-th, $\mu+1$-th, $1$-th row replaced by,
\[
\begin{aligned}
\left(
\begin{array}{cccc|ccc}
0&\cdots&0&-\alpha&0&\cdots&0
\end{array}
\right),\qquad
\left(
\begin{array}{cccc|ccc}
0&\cdots&0&-\frac{\alpha}{a}&0&\cdots&0
\end{array}
\right),\\[2mm]
\left(
\begin{array}{cccc|ccc}
-\frac{a}{(\mu-1)!}&-\frac{a}{(\mu-2)!}&-\cdots&-\frac{a}{0!}&0&\cdots&0
\end{array}
\right),\quad\mbox{respectively.}
\end{aligned}
\]
Now multiply the matrix in (\ref{hattau}) from the left with the matrix
\[
\sum_{1\le j\le i\le 2\mu} \frac{(-1)^{i-j}}{(i-j)!}E_{ij},
\]
Then $\hat\tau(s)$ does not change and now becomes equal to
\begin{equation}
\label{hattau2}
\hat\tau(s)=\det \sum_{i=1}^\mu E_{\mu+i,\mu+i}
+\sum_{j=1}^{2\mu}\left((T_{2\mu-j}^\mu(s)\right)^{(\mu-i)}E_{ij}
,
\end{equation}
with
\begin{equation}
\label{T}
T^\mu_{k}(s)=\sum_{j=0}^{k-\mu}
\frac{(-1)^j}{j!}S_{k-j}(s)-a^2\sum_{j=k-\mu+1}^{k}
\frac{(-1)^j}{j!}S_{k-j}(s)\quad\mbox{and }
\left( T^\mu_{k}(s)\right)^{(p)}=\frac{\partial^p T^\mu_{k}(s)}{\partial
s_1^p}.
\end{equation}

Multiplying the determinant of the other $\hat\tau_{ij}(s)$ by the same matrix,
 one obtains that
$\hat\tau_{12}(s)$, $\hat\tau_{13}(s)$, $\hat\tau_{32}(s)$, respectively
 is  the same determinant with the $1$-th, $\mu+1$-th, $1$-th row replaced by,
\[
\begin{aligned}
 &\alpha\left(
\begin{array}{cccc|ccc}
\frac{(-1)^\mu}{(\mu-1)!}&\frac{(-1)^{\mu-1}}{(\mu-2)!}&\cdots&
\frac{-1}{0!}&0&\cdots&0
\end{array}
\right),\qquad
\frac{\alpha}{a}\left(
\begin{array}{cccc|ccc}
\frac{(-1)^\mu}{(\mu-1)!}&\frac{(-1)^{\mu-1}}{(\mu-2)!}&\cdots&
\frac{-1}{0!}&0&\cdots&0
\end{array}
\right),\\[2mm]
 &\qquad\qquad\qquad\qquad -a\left(
\begin{array}{cccc|ccc}
0&\cdots&0&1&0&\cdots&0
\end{array}
\right),\quad \mbox{respectively.}
\end{aligned}
\]
Now permuting the first $\mu$ rows of the matrix gives that
\begin{equation}
\label{tT}
\begin{aligned}
\hat\tau(s)=&(-)^{\frac{\mu(\mu-1)}{2}}\mbox{W}\left(
T^\mu_{2\mu-1}( s  ),  T^\mu_{2\mu-2}( s  ),
\cdots ,   T^\mu_{\mu}( s  )
\right),
\\[2mm]
\hat\tau_{12}(s)=&-(-)^{\frac{\mu(\mu-1)}{2}}\alpha \mbox{W}\left(
T^\mu_{2\mu-1}( s  )+\frac{T^\mu_{2\mu-2}( s  )}{\mu-1},  T^\mu_{2\mu-2}( s
)+\frac{T^\mu_{2\mu-3}( s  )}{\mu-2},
\cdots ,   T^\mu_{\mu+1}( s  )+ T^\mu_{\mu}( s  )
\right),
\\[2mm]
\hat\tau_{13}(s)=&-(-)^{\frac{\mu(\mu-1)}{2}}\frac{\alpha }{a}
\mbox{W}\left(
T^\mu_{2\mu-1}( s  )+\frac{T^\mu_{2\mu-2}( s  )}{\mu-1},  T^\mu_{2\mu-2}( s
)+\frac{T^\mu_{2\mu-3}( s  )}{\mu-2},
\cdots ,   T^\mu_{\mu+1}( s  )+ T^\mu_{\mu}( s  ),T^\mu_{\mu-1}( s  )
\right),
\\[2mm]
\hat\tau_{32}(s)=&(-)^{\frac{\mu(\mu-1)}{2}}a\mbox{W}\left(
T^\mu_{2\mu-1}( s  ),  T^\mu_{2\mu-2}( s  ),
\cdots ,   T^\mu_{\mu+1}( s  )
\right),
\end{aligned}
\end{equation}
where W stands for the Wronskian determinant:
\[
W(f_1(s),f_2(s), \ldots, f_n(s))
=\det \left(\frac{\partial^{i-1}f_j(s)}{\partial s_1^{i-1}}\right)_{1\le i,j\le
n}.
\]
Thus, by (\ref{bt}) we have an expression for $\beta_{ij}(s)$ and hence can
calculate the
$\omega_i(s)$'s.
Now put all $s_j=0$ for $j>1$ and write $x$ for $s_1$, then
\begin{equation}
\label{ooo}
\begin{aligned}
\omega_1(x)=&-{a}(1-x)\frac{\mbox{W}\left(
T^\mu_{2\mu-1}( x  ),  T^\mu_{2\mu-2}( x  ),
\cdots ,   T^\mu_{\mu+1}( x  )
\right)}{\mbox{W}\left(
T^\mu_{2\mu-1}( x  ),  T^\mu_{2\mu-2}( x  ),
\cdots ,   T^\mu_{\mu}( x  )
\right)},\\[3mm]
\omega_2(x)=&-\frac{\alpha }{a}\frac{\mbox{W}\left(
T^\mu_{2\mu-1}( x  )+\frac{T^\mu_{2\mu-2}( x  )}{\mu-1},  T^\mu_{2\mu-2}( x  )+
\frac{T^\mu_{2\mu-3}( x  )}{\mu-2},
\cdots ,   T^\mu_{\mu+1}( x  )+ T^\mu_{\mu}( x  ),T^\mu_{\mu-1}( x  )
\right)}{\mbox{W}\left(
T^\mu_{2\mu-1}( x  ),  T^\mu_{2\mu-2}( x  ),
\cdots ,   T^\mu_{\mu}( x  )
\right)},\\[3mm]
\omega_3(x)=&-\alpha x\frac{\mbox{W}\left(
T^\mu_{2\mu-1}( x  )+\frac{T^\mu_{2\mu-2}( x  )}{\mu-1},  T^\mu_{2\mu-2}( x  )+
\frac{T^\mu_{2\mu-3}( x  )}{\mu-2},
\cdots ,   T^\mu_{\mu+1}( x  )+ T^\mu_{\mu}( x  )
\right)}{\mbox{W}\left(
T^\mu_{2\mu-1}( x  ),  T^\mu_{2\mu-2}( x  ),
\cdots ,   T^\mu_{\mu}( x  )
\right)}
\end{aligned}
\end{equation}
satisfy the Euler top equations (\ref{eq:euta}).
We will show later that
$
\sum_{i=1}^3 \omega_i(x)=-\mu^2
$.

Note, see (\ref{tT}), that  $\hat\tau(s)$ and $\hat\tau_{ij}(s)$ are
Wronskians of functions wich satisfy
\[
\frac{\partial f(s)}{\partial s_p}=\frac{\partial^p f(s)}{\partial s_1^p}\qquad
p=2,3,4,\ldots.
\]
Hence they are  1-component KP tau-functions.
In the next sections we will show that these Wronskians can be obtained in the
(1-component) 2-vector 1-constrained CKP hierarchy.

\section{The 2-vector 1-constrained CKP hierarchy}
\label{Sec4}
The Lax operator $L$ of the (1-component) 2-vector 1-constrained CKP hierarchy
can be written as (see \cite{AL2})
\begin{equation}
\label{laxCKP}
L=\partial  +\Phi_1(t)\partial ^{-1}\Phi_1^*( t)+\Phi_2(t)\partial
^{-1}\Phi_2^*( t),
\end{equation}
where $\Phi_j(t)$ is an eigenfunction and $\Phi_j^*(t)=\Phi_j(\tilde t)$ an
adjoint eigenfunction, satisfying
\begin{equation}
\label{eigen}
\frac{\partial \Phi_j(t)}{\partial t_n}=(L^n)_+\Phi_j(t),\qquad \frac{\partial
\Phi_j^*(t)}{\partial
t_n}=-((L^*)^n)_+\Phi_j^*(t).
\end{equation}
Recall that the Sato KP Grassmannian
consists of all linear spaces
\[
W\subset
H_+\oplus H_-={\mathbb C}[z]\oplus z^{-1}{\mathbb C}[[z^{-1}]],
\]
such that the projection on $H_+$ has  finite index. We introduce a natural
filtration on
 Grassmannian
\[
\cdots\subset H_{k+1}\subset H_k\subset H_{k-1}\subset H_{k-2} \subset \cdots ,
\]
consisting of the linear subspaces
\[
H_k=\{ \sum^{N}_{j=k} a_j
z^{j} | a_j \in \mathbb{C}\}.
\]
On the space $H$ we have a bilinear form, viz.  if
$f(z)=\sum_j a_j z^j$ and $g(z)=\sum_j b_j z^j$ are in
$H$, then we define
\begin{equation}
\label{hel1}
(f(z),g(z))=Res_{z}f(z)g(z)= \sum_j a_jb_{-j-1}.
\end{equation}
Then the polynomial Sato Grassmannian $Gr (H)$ consists
of all linear subspaces of
$W\subset H$ such that
\begin{equation}
\label{kl}
H_{k}\subset W\subset H_\ell\qquad\text{ for certain }
 k>\ell.
\end{equation}
The space $Gr (H)$ has a subdivision into different components:
\[
Gr^{(j)} (H) = \{ W \in Gr (H)| H_k\subset W,\ j=k-\dim
(W/H_k)\}.
\]
Clearly, the subspace $H_k$ belongs to $Gr^{(k)} (H)$.
The polynomial CKP Sato Grassmannian consists of linear subspaces of $Gr^{(0)}
(H)$
such that  for any
$f(z),g(z)\in W$ one has $(f(z),g(-z))=0$.
To describe the spaces corresponding to the  2-vector 1-constrained CKP
hierarchy, such $W$ must also satisfy the following
condition \cite{HL1}, \cite{HL2}, \cite{AL2}. There exists a subspace
\begin{equation}
W'\subset W\quad \mbox{of codimension 2 such that }zW'\subset W.
\end{equation}
We assume that there is no larger subspace $W'$ with $zW'\subset W$.
Let $\psi_W(t,z)$ be the  wave function corresponding to such $W$, then the
$\Phi_j(t)$ can be constructed as follows.
Let
\[
zW+W=W\oplus \mathbb{C}zf_1(z)\oplus \mathbb{C}zf_2(z)
\]
with $f_i(z)\in W$. Choose two independent elements $h_i(z)\in
\mathbb{C}f_1(z)\oplus \mathbb{C}f_2(z)$ such that
\[
(h_1(z),zh_2(-z)=(h_2(z),zh_1(-z))=0,
\]
then up to a scalar constant $c_j$ one has
\[
\Phi_j(t)=c_j\left( \psi_W(t,z), zh_j(-z)\right).
\]

\section{B\"acklund--Darboux transformations}
\label{SecBD}

In the next section we will define subspaces $W$ that are related to the
tau-functions $\hat\tau(s)$ of Section
\ref{Sec3}. Since
B\"acklund--Darboux transformations will play an important role in our
construction, we will describe the elementary ones
first.
For $W\in Gr(H)$, let $W^{\perp}$ be the orthocomplement of $W$ in
$H$ w.r.t. the bilinear form (\ref{hel1}). Then,    $W^{\perp}$ also belongs
to $Gr(H)$.

For
each $W \in Gr(H)$ we denote the wave function corresponding to $W$ by
$\psi_W$. The dual wave function of $\psi_W$, which we denote by
$\psi_W^*$ can be characterized as follows \cite{Sh},
\cite{HL}:
\begin{proposition}
\label{C1}
Let $W$ and $\tilde{W}$ be two subpaces in $Gr(H)$. Then $\tilde{W}$ is the
space $W^*$
 corresponding to the dual wave function, if and only if
$\tilde{W}=W^{\perp}$ with
$W^{\perp}$ the orthocomplement of $W$ w.r.t. the bilinear form (\ref{hel1}) on
$H$. Moreover
\[
\left(\psi_W(t,z),\psi_W^*(s,z)\right)=0.
\]
\end{proposition}
Let $W\in Gr^{(k)}(H)$ then
\[
\psi_W(t,z)=P_W(t,\partial)e^{\sum_{j=1}^\infty t_jz^j},
\qquad
\psi_W^*(t,z)=P_W^{*-1}(t,\partial)e^{-\sum_{j=1}^\infty t_jz^j},
\]
where  $P_W(t,\partial)$ is an $k^{th}$ order pseudo-differential operator.
The corresponding KP Lax operator $ L_W$ is equal to
\begin{equation}
\label{g0}
 L_W(t,\partial)=P_W(t,\partial)\partial P_W^{-1}(t,\partial).
\end{equation}
{}{}From now on we will use the notation $\psi_W$ and $ L_W$  whenever we want
to emphasize its dependence on a
point $W$ of the Sato Grassmannian $Gr(H)$.

Eigenfunctions $\Phi(t)$ and adjoint eigenfunctions $\Psi(t)$ of the KP Lax
operator, satisfy
(\ref{eigen}) and  can be expressed in wave and adjoint wave functions, viz.
there
exist functions $f,g\in H$ such that
\begin{equation}
\label{g1}
\Phi(t)=\left(\psi_W(t,z),f(z)\right), \qquad
\Psi(t)=\left(\psi_W^*(t,z),g(z)\right).
\end{equation}
Such (adjoint) eigenfunctions induce elementary B\"acklund--Darboux
transformations \cite{HL}.
Assume that we have the following data $W\in Gr^{(k)}(H)$, $W^\perp$,
$\psi_W(t,z)$
and $\psi_W^*(t,z)$, then the (adjoint) eigenfunctions (\ref{g1})
induce new KP wave functions:
\begin{equation}
\label{g2}
\begin{aligned}
\psi_{W'}(t,z)&=\left(\Phi(t)\partial
\Phi(t)^{-1}\right)\psi_{W}(t,z),&\quad
\psi_{W'}^*(t,z)&=\left(\Phi(t)
\partial\Phi(t)^{-1}\right)^{*-1}\psi_{W}^*(t,z),\\
\psi_{{W''}}(t,z)&=\left(-\Psi(t)\partial
\Psi(t)^{-1}\right)^{*-1}\psi_{W}(t,z),&\quad
\psi_{{W''}}^*(t,z)&=\left(-\Psi(t)\partial
\Psi(t)^{-1}\right)\psi_{W}^*(t,z),
\end{aligned}
\end{equation}
and new tau-functions
\begin{equation}
\label{g22}
\tau_{W'}(t)=\Phi(t)\tau_w(t),\qquad\tau_{{W''}}(t)=\Psi(t)\tau_W(t),
\end{equation}
where
\begin{equation}
\label{g3}
\begin{aligned}
W'&=\{ w\in W|(w(z),f(z))=0\}\in Gr^{(k+1)}(H),\quad
{W'}^\perp=W^\perp + \mathbb{C}f,\\
{W''}&=W+\mathbb{C}g\in Gr^{(k-1)}(H),\qquad {{W''}}^\perp =
\{ w\in W^\perp|(w(z),g(z))=0\}.
\end{aligned}
\end{equation}
Now applying $n$ consecutive elementary B\"acklund--Darboux transformations
such that one obtains
\[
W'=\{ w\in W|(w(z),f_i(z))=0, \ i=1,2,\ldots, n\}
\]
from $W$, then (see \cite{HL})
\[
\tau_{W'}(t)=W(\Phi_1(t),\Phi_2(t), \ldots, \Phi_n(t))\tau_W(t),
\]
where one has to take derivatives w.r.t. $x$ and where
\[
\Phi_j(t)=(\psi_W(t,z),f_j(z))
\]
and
\begin{equation}
\label{ppp}
\psi_{W'}(t,z)=\frac{1}{\tau_{W'}(t)}W\left(\Phi_1(t),\Phi_2(t), \ldots,
\Phi_n(t),
\psi_W(t,z)\right)
{}.
\end{equation}

\section{Subspaces $W_\mu$}

In this section we will construct Subspaces $W_\mu$ in the 2-vector
1-constrained CKP hierarchy
related to the solutions of Section \ref{Sec3} of the time-dependent Euler top
equations.
Let $a\in\mathbb{C}$ with $a\ne 0,\pm i$ be the parameter of Section
\ref{Sec3}. Define $b=-a^2$, then $b\ne 0,1$
 and introduce
\begin{equation}
\label{r_0}
r_0(z)=b e^{z}.
\end{equation}
{\it Unfortunately $r_0(z)$  is not an element of $H$. However, since we always
assume that $H_k\subset W$
for $k>>0$, we will write $e^z$ and will mean in fact $\sum_{j=0}^N
\frac{z^j}{j!}$ with $N>2k>>0$.}
Having this in mind, we  define for $i=1,2,\cdots$ the elements.
\begin{equation}
\label{r}
r_i(z)=z^{-i} \left( b e^{z}+(1-b)\sum_{j=0}^{i-1} \frac{z^j}{j!}\right).
\end{equation}
Note that
\begin{equation}
\label{ri+1}
r_{i+1}(z)=z^{-1}\left( r_i(z)+\frac{1-b}{i!}\right)
\end{equation}
and a straightforward calculation shows:
\begin{lemma}
\label{L2.1}
For $i,j>0$
\[
(r_i(z),r_j(-z))=0.
\]
\end{lemma}
\vskip 10pt
\noindent
Now define for $\mu=1,2,\cdots$, the point $W_\mu\in Gr(H)$
\begin{equation}
\label{wmu}
W_\mu=\text{linear span}\, \{ r_1(z), r_2(z), \ldots,r_\mu(z)\}\oplus
H_{\mu}.
\end{equation}
{}From now on we will assume that $\mu$ can also be 0, then
$
W_0=H_0
$.
{}From the definition (\ref{r}) of the functions $r_i(z)$ it is clear that
\[
(f(z),r_i(-z))= (f(z),g(-z))=0\quad \text{for all } f(z),g(z)\in H_{\mu},\quad
0\le i\le\mu
{}.
\]
{}From Lemma \ref{L2.1} it is then clear that $W_\mu$ satisfies the CKP
condition, to be more
precise
\begin{proposition}
\label{prop2.1}
$W_\mu\in Gr^{(0)}(H)$ satisfies the CKP condition and
\[
W_\mu=\{ f(z)\in H_{-\mu}|\, (f(z),r_i(-z))=0, \ \text{for } 1\le i\le \mu\}.
\]
\end{proposition}
\vskip 10pt\noindent
Next define  the subspace
$U_\mu\subset W_\mu$ of codimension 2 for $\mu\ge 2$, of codimension 1 if
$\mu=1$ and of
codimension 0 if $\mu=0$ by
\begin{equation}
\label{umu}
U_\mu=\{ f(z)\in W_\mu |\, (f(z),1)=(f(z), r_0(-z))=0\}.
\end{equation}
Now let $g(z)\in U_\mu$, then $zg(z)\in H_{-\mu+1}$ and
$
(zg(z),r_j(-z))=0$ for all $1\le j\le\mu$.

This follows from the following observation:
\[
(zg(z),r_j(-z))=(g(z),
zr_j(-z))=\left(g(z),-r_{j-1}(-z)-\frac{1-b}{(j-1)!}\right)=0,
\]
for $j=1,2,\ldots,\mu$,
since $g(z)$ is perpendicular to $1$, $r_i(-z)$ for $ 0\le i\le \mu$.
Hence, $W_\mu$ has a subspace $W'$ of codimension 2 such that
$zW'\subset W_\mu$, hence
\begin{proposition}
\label{prop2.2}
$W_\mu$ with $\mu>1$ also belongs to the 2-vector 1-constrained KP hierarchy.
\end{proposition}
\vskip 10pt\noindent
Note that $W_1$ belongs to the 1-vector 1-constrained KP.
{}From Proposition \ref{prop2.1}  and Section \ref{SecBD} it is clear that
$W_\mu$ can be obtained from $H_{-\mu}\in
Gr^{(\mu)}(H)$ by $\mu$ consecutive elementary B\"acklund--Darboux
transformations.
Now $\tau_{H_{-\mu}}=1$ and $\psi_{H_{-\mu}}(t,z)=z^{-\mu}\psi_0(t,z)$ where
$\psi_0(t,z)=e^{\sum_{i=0}^\infty t_iz^i}$.
Let $\tau_\mu(t)=\tau_{W_{\mu}}(t)$ and $\psi_\mu(t,z)=\psi_{W_\mu}(t,z)$,
then
\begin{equation}
\label{taumu}
\tau_\mu(t)=W\left(R_1^\mu(t),R_2^\mu(t),\ldots,R_\mu^\mu(t)\right),
\end{equation}
where
\begin{equation}
\label{R}
\begin{aligned}
R_i^\mu(t):=&(z^{-\mu}\psi_0(t,z),r_i(-z))
=(\psi_0(t,z),z^{-\mu}r_i(-z))\\[2mm]
=&\sum_{k=0}^{i-1} \frac{(-1)^{k-i}}{k!}S_{\mu +i-k-1}(t)
+b\sum_{k=i}^{\mu+i-1} \frac{(-1)^{k-i}}{k!}S_{\mu +i-k-1}(t).
\end{aligned}
\end{equation}
Here
$S_k(t)$ are the elementary Schur functions.
The corresponding wave function is given by
\begin{equation}
\label{psimu}
\begin{aligned}
\psi_\mu(t,z)
=\frac{1}{\tau_\mu(t)}
W
\left(
(R_1^\mu(t),R_2^\mu(t),\ldots,R_\mu^\mu(t), z^{-\mu}\psi_0(t,z)
\right ).
\end{aligned}
\end{equation}
Note that
\[
R_i^\mu(t)=(-)^i T_{\mu+i-1}^\mu (t),\qquad\mbox{with }b=-a^2,
\]
Hence
\[
\tau_\mu(s)=(-)^{\frac{\mu(\mu +1)}{2}}\hat\tau(s).
\]
In order to describe the other tau-functions of Section \ref{Sec3}, we
 want to find the right expression for the Lax operator
$L=L_\mu=L_{W_\mu}$. For this we study $W_\mu$  and $zW_\mu$. Recall from
(\ref{wmu}) that
\[
W_\mu=\text{linear span}\, \{ r_1(z), r_2(z), \ldots,r_\mu(z)\}\oplus
H_{\mu},
\]
and
\[
W_\mu^\perp=\text{linear span}\, \{ r_1(-z), r_2(-z), \ldots,r_\mu(-z)\}\oplus
H_{\mu},
\]
hence
\[
\begin{aligned}
zW_\mu=&\text{linear span}\, \{ zr_1(z), zr_2(z), \ldots,zr_\mu(z)\}\oplus
H_{\mu +1},\\[2mm]
(zW_\mu)^\perp=&\text{linear span}\, \{ z^{-1}r_1(-z), z^{-1}r_2(-z),
\ldots,z^{-1}r_\mu(-z)\}\oplus
H_{\mu -1}.
\end{aligned}
\]
{\it From now on we assume in this section
that $\mu>1$.} In that case
it is straightforward to check
that
\[
zW_\mu +W_\mu=W_\mu\oplus \mathbb{C}zr_1(z)\oplus \mathbb{C}zr_2(z).
\]
Putting
\begin{equation}
h_1(z)=r_1(z)-r_2(z)\quad\text{and}\quad h_2(z)=r_2(z),
\end{equation}
one easily verifies that
\begin{equation}
\label{hh}
(h_1(z),zh_2(-z))=(h_2(z),zh_1(-z))=(h_1(-z),zh_2(z))=(h_2(-z),zh_1(z))=0.
\end{equation}
Using the construction of the Lax operator as in Section \ref{Sec4} we see that
\begin{equation}
\label{l1}
\begin{aligned}
L_\mu=&\partial+\sum_{i=1}^2 c_i (\psi_\mu(t,z),zh_i(-z))\partial^{-1}
(\psi_\mu^*(t,z),zh_i(z))\\[3mm]
=&\partial+\sum_{i=1}^2 c_i (\psi_\mu(t,z),zh_i(-z))\partial^{-1}
(\psi_\mu(\tilde t,z),zh_i(-z)).
\end{aligned}
\end{equation}
We want to determine the $c_i$'s, for this we let $L_\mu$ act on $\psi_\mu$,
this gives
\begin{equation}
\label{l2}
\begin{aligned}
z\psi_\mu(t,z)=&\frac{\partial \psi_\mu(t,z)}{\partial x}+\sum_{i=1}^2
c_i (\psi_\mu(t,z),zh_i(-z))\partial^{-1}
(\psi_\mu^*( t,z),zh_i(z))\psi_\mu(t,z)\\[3mm]
=&\frac{\partial \psi_\mu(t,z)}{\partial x}+\sum_{i=1}^2 c_i
(\psi_\mu(t,z),zh_i(-z))(\psi_\mu(\tilde t,z),zh_i(-z))
\psi_{W_\mu+\mathbb{C}zh_i(z)}(t,z).
\end{aligned}
\end{equation}
Now take the bilinear form with the elements
$h_j(-z)$. Since (\ref{hh}) holds,
and $h_1(-z)$ (resp. $h_2(-z)$) is perpendicular to $W_\mu$ and
$W_\mu+\mathbb{C}zh_2(z)$ (resp. $W_\mu+\mathbb{C}zh_1(z)$)
we obtain
\[
(\psi_\mu(t,z), zh_i(-z))=c_i(\psi_\mu(t,z),zh_i(-z))(\psi_\mu(\tilde
t,z),zh_i(-z))
(\psi_{W_\mu+\mathbb{C}zh_i(z)}(t,z),h_i(-z)).
\]
Hence
\begin{equation}
\label{ci}
c_i=\left( (\psi_\mu(\tilde t,z),zh_i(-z))
\left(\psi_{W_\mu+\mathbb{C}zh_i(z)}(t,z),h_i(-z)\right)\right)^{-1}.
\end{equation}
We are now going to determine these $c_i$'s. Note that
\begin{equation}
\label{hhhh}
zh_1(z)=r_0(z)-r_1(z)\quad \text{and}\quad zh_2(z)=1-b+r_1(z).
\end{equation}
Using this we see
that
\[
\begin{aligned}
W_\mu+\mathbb{C}zh_1(z)=&\text{linear span}\, \{ r_0(z),r_1(z),\ldots
r_\mu(z)\}+H_\mu,\\[2mm]
W_\mu+\mathbb{C}zh_2(z)=&\text{linear span}\, \{ 1,r_1(z),r_2(z),\ldots
r_\mu(z)\}+H_\mu.
\end{aligned}
\]
The fact that
\[
(r_0(z),r_1(-z))=-b,\ (r_0(z),r_i(-z))=0 \ \text{and}\
(1,r_j(-z))=-\frac{1}{(j-1)!}\quad\text{for } i>1,\ j\ge 1,
\]
gives the following, more convenient description of $W_\mu+\mathbb{C}zh_1(z)$
and
$W_\mu+\mathbb{C}zh_2(z)$:
\begin{equation}
\label{ww}
\begin{aligned}
W_\mu+\mathbb{C}zh_1(z)=&\{ f(z)\in H_{-\mu}|\, (f(z),r_i(-z))=0\
\text{for } i=2,3,\ldots,\mu \},\\[2mm]
W_\mu+\mathbb{C}zh_2(z)=&\{ f(z)\in H_{-\mu}|\,
\left(f(z),r_{i+1}(-z)-\frac{r_i(-z)}{i}\right)=0\
\text{for } i=1,2 ,\ldots,\mu-1 \}.
\end{aligned}
\end{equation}
Thus,
\begin{equation}
\label{tauww}
\begin{aligned}
\tau_{W_\mu+\mathbb{C}zh_1(z)}(t)=&
\det\left(
\left(  \psi_0(t,z),z^{i-\mu-1}r_{j+1}(-z)  \right)
\right)_{1\le i,j\le\mu-1}\\[3mm]
=&W(R^\mu_2(t),R^\mu_3(t), \ldots, R^\mu_\mu (t)),\\[3mm]
\tau_{W_\mu+\mathbb{C}zh_2(z)}(t)=&
\det\left(
\left(  \psi_0(t,z),z^{i-\mu-1}\left(r_{j+1}(-z) -\frac{r_j(-z)}{j}\right)
\right)
\right)_{1\le i,j\le\mu-1}\\[3mm]
=&W\left(R^\mu_2(t)-R^\mu_1(t),R^\mu_3(t)-\frac{R^\mu_2(t)}{2},
\ldots, R^\mu_\mu (t)-\frac{R^\mu_{\mu-1}(t)}{\mu-1}\right).
\end{aligned}
\end{equation}
and the corresponding wave functions are equal to:
\begin{equation}
\label{psiw1}
\psi_{W_\mu+\mathbb{C}zh_1(z)}(t,z)=\frac{1}{\tau_{W_\mu+\mathbb{C}zh_1(z)}(t)}
W
\left (
W(R^\mu_2(t),R^\mu_3(t), \ldots, R^\mu_\mu (t), z^{-\mu}\psi_0(t,z)
\right ).
\end{equation}
and
\begin{equation}
\label{psiw2}
\begin{aligned}
&\psi_{W_\mu+\mathbb{C}zh_2(z)}(t,z)
=\frac{1}{\tau_{W_\mu+\mathbb{C}zh_2(z)}(t)}\times \\[2mm]
&
W
\left (
R^\mu_2(t)-R^\mu_1(t),R^\mu_3(t)-\frac{R^\mu_2(t)}{2},
\ldots, R^\mu_\mu (t)-\frac{R^\mu_{\mu-1}(t)}{\mu-1}
, z^{-\mu}\psi_0(t,z)
\right ).
\end{aligned}
\end{equation}
{}From this we deduce that
\begin{equation}
\label{tt}
\begin{aligned}
\left(\psi_{W_\mu+\mathbb{C}zh_1(z)}(t,z),h_1(-z)\right)=&
\left(\psi_{W_\mu+\mathbb{C}zh_1(z)}(t,z),r_1(-z)\right)
=(-)^{\mu-1}\frac{\tau_\mu(t)}{\tau_{W_\mu+\mathbb{C}zh_1(z)}(t)},\\[3mm]
\left(\psi_{W_\mu+\mathbb{C}zh_2(z)}(t,z),h_2(-z)\right)=&
\left(\psi_{W_\mu+\mathbb{C}zh_2(z)}(t,z),r_2(-z)\right)
=(-)^{\mu-1}\frac{\tau_\mu(t)}{\tau_{W_\mu+\mathbb{C}zh_2(z)}(t)}.
\end{aligned}
\end{equation}
For the other eigenfunctions we find, using (\ref{hhhh}):
\begin{equation}
\label{tauw'1}
\begin{aligned}
\left(\psi_\mu(\tilde t,z),zh_1(-z)\right)&=(-)^{\mu+1}
\frac{\tau_{W'}(\tilde t)}{\tau_{\mu}(\tilde t)},\\[2mm]
\left(\psi_\mu(\tilde t,z),zh_2(-z)\right)&=(b-1)
\frac{\tau_{W''}(\tilde t)}{\tau_{\mu}(\tilde t)},
\end{aligned}
\end{equation}
where
\begin{equation}
\label{w'}
\begin{aligned}
W'=&\{ f(z)\in H_{-\mu}|\, (f(z),r_i(-z))=0\
\text{for } i=0,1,\ldots,\mu \},\\[2mm]
W''=&\{ f(z)\in H_{-\mu}|\,(f(z),1)=0\ \text{and }
(f(z),r_i(-z))=0\
\text{for } i=1,2,\ldots,\mu \}
\end{aligned}
\end{equation}
and
\begin{equation}
\label{tauw'2}
\begin{aligned}
\tau_{W'}(t)
=&W\left(R_0^\mu(t),R_1^\mu(t),R_2^\mu(t),\ldots,R_\mu^\mu(t)\right),\\[2mm]
\tau_{W''}(t)=
&W\left(R_1^\mu(t),R_2^\mu(t),\ldots,R_\mu^\mu(t),S_{\mu-1}(t)\right).
\end{aligned}
\end{equation}
Now combining (\ref{ci}), (\ref{tauw'1}) and (\ref{tt}), we find that
\begin{equation}
\label{ci1}
\begin{aligned}
c_1=&\frac{\tau_\mu(\tilde t)\tau_{W_\mu+\mathbb{C}zh_1(z)}(t)}
{\tau_{W'}(\tilde t)\tau_\mu(t)}=\frac{\tau_{W_\mu+\mathbb{C}zh_1(z)}(t)}
{\tau_{W'}(\tilde t)},\\[3mm]
c_2=&(-)^{\mu-1}(b-1)^{-1}\frac{\tau_\mu(\tilde
t)\tau_{W_\mu+\mathbb{C}zh_2(z)}(t)}
{\tau_{W''}(\tilde
t)\tau_\mu(t)}=(-)^{\mu-1}(b-1)^{-1}\frac{\tau_{W_\mu+\mathbb{C}zh_2(z)}(t)}
{\tau_{W''}(\tilde t)},
\end{aligned}\end{equation}
since $\tau_\mu(\tilde t)=\tau_\mu( t)$.
Since these $c_i$'s are just constants, it suffices to substitute $t=0$, i.e.
$t_j=0$
for all $j=1,2,3,\ldots$, in (\ref{ci1}), this gives
\begin{equation}
\label{ci2}
c_1=\frac{\tau_{W_\mu+\mathbb{C}zh_1(z)}(0)}
{\tau_{W'}(0)},\qquad
c_2=(-)^{\mu-1} (b-1)^{-1}\frac{\tau_{W_\mu+\mathbb{C}zh_2(z)}(0)}
{\tau_{W''}(0)}.
\end{equation}
We now calculate these tau-functions for $t=0$:
\begin{equation}
\label{t1(0)}
\begin{aligned}
\tau_{W_\mu+\mathbb{C}zh_1(z)}(0)=&
\det\left(
\left(  z^{i-\mu-1},r_{j+1}(-z)  \right)
\right)_{1\le i,j\le\mu-1}\\[2mm]
=&
\det\left(
\frac{(-)^{\mu-i}b}{(\mu+j-i+1)!}
\right)_{1\le i,j\le\mu-1}\\[2mm]
=&(-)^{\frac{\mu(\mu-1)}{2}}b^{\mu-1}
\det\left(
\frac{1}{(\mu+j-i+1)!}
\right)_{1\le i,j\le\mu-1}\\[2mm]
=&(-)^{\frac{\mu(\mu-1)}{2}}
b^{\mu-1}S_{\mu+1,\mu+1,\ldots,\mu+1}^{(\mu-1)}
(1,0,0,\ldots),
\end{aligned}
\end{equation}
where (see \cite{Mac}
\[
S_{\lambda_1,\lambda_2,\ldots,\lambda_k}^{(k)}(t_1,t_2,t_3,\ldots)=
\det \left(S_{\lambda_i+j-i}(t_1,t_2,t_3,\ldots)\right)_{1\le i,j\le k},
\]
the Schur function corresponding to the partition
$\lambda_1,\lambda_2,\ldots,\lambda_k$.
Here $S_\ell(t_1,t_2,t_3,\ldots)$  is the elementary Schur function.
In a similar way one shows that
\begin{equation}
\label{t2(0)}
\begin{aligned}
\tau_{W'}(0)=&\det\left(\left(z^{i-\mu-1},r_{j-1}(-z)\right)\right)_{1\le
i,j\le \mu+1}\\[2mm]
=&(-)^{\frac{\mu(\mu-1)}{2}+1}
b^{\mu}S_{\mu-1,\mu-1,\ldots,\mu-1}^{(\mu+1)}(1,0,0,\ldots)
\end{aligned}
\end{equation}
and
\begin{equation}
\label{t3(0)}
\begin{aligned}
\tau_{W''}(0)=&
\det
\left (
\begin{array}{cccc}
(z^{-\mu},r_1(-z))
&\cdots&(z^{-\mu},r_\mu(-z))& (z^{-\mu},1)\\[2mm]
(z^{1-\mu},r_1(-z))
&\cdots&(z^{1-\mu},r_\mu(-z))& (z^{1-\mu},1)\\[2mm]
\vdots&&\vdots&\vdots\\[2mm]
(r_1(-z),1)
&\cdots&(r_\mu(-z),1)&(1,1)
\end{array}
\right )\\[2mm]
=&(-)^{\frac{\mu(\mu-1)}{2}}b^{\mu-1}S_{\mu,\mu,\ldots,\mu,
\mu-1}^{(\mu)}(1,0,0,\ldots).
\end{aligned}
\end{equation}
And finally the most complicated one:
\begin{equation}
\label{t4(0)}
\begin{aligned}
\tau_{W_\mu+\mathbb{C}zh_2(z)}(0)=&
\det\left(
\left(  z^{i-\mu-1},\left(r_{j+1}(-z) -\frac{r_j(-z)}{j}\right) \right)
\right)_{1\le i,j\le\mu-1}\\[2mm]
=&
\det\left(
\frac{(-)^{\mu-i}b}{(\mu+j-i+1)!}-\frac{(-)^{\mu-i}b}{(\mu+j-i)! j}
\right)_{1\le i,j\le\mu-1}\\[2mm]
=&
(-)^{\frac{(\mu-1)(\mu+2)}{2}}b^{\mu-1}\det\left(\frac{\mu-i+1}{(\mu+j-i+1)!
j}\right)_{1\le i,j\le\mu-1}\\[2mm]
=&(-)^{\frac{(\mu-1)(\mu+2)}{2}}\mu b^{\mu-1}
\det\left(\frac{1}{(\mu+j-i+1)! }\right)_{1\le i,j\le\mu-1}\\[2mm]
=&(-)^{\frac{(\mu-1)(\mu+2)}{2}}\mu b^{\mu-1}
S_{\mu+1,\mu+1,\ldots,\mu+1
}^{(\mu-1)}(1,0,0,\ldots).
\end{aligned}
\end{equation}
We conclude from all this that
\begin{equation}
\label{ci3}
c_1=-b^{-1}\frac{S_{\mu+1,\mu+1,\ldots,\mu+1}^{(\mu-1)}(1,0,0,\ldots)}
{S_{\mu-1,\mu-1,\ldots,\mu-1}^{(\mu+1)}(1,0,0,\ldots)},\quad
c_2=(b-1)^{-1}\mu \frac{S_{\mu+1,\mu+1,\ldots,\mu+1
}^{(\mu-1)}(1,0,0,\ldots)}{S_{\mu,\mu,\ldots,\mu,
\mu-1}^{(\mu)}(1,0,0,\ldots)}.
\end{equation}
Now using the fact that
\[
\begin{aligned}
S_{\mu+1,\mu+1,\ldots,\mu+1}^{(\mu-1)}(1,0,0,\ldots)=&
S_{\mu-1,\mu-1,\ldots,\mu-1}^{(\mu+1)}(1,0,0,\ldots)
=\mu \frac{\prod_{i=1}^{\mu-1} (i!)^2}{\prod_{i=1}^{2\mu-1} i!},\\[3mm]
S_{\mu,\mu,\ldots,\mu,
\mu-1}^{(\mu)}(1,0,0,\ldots)=&\mu^2
\frac{\prod_{i=1}^{\mu-1} (i!)^2}{\prod_{i=1}^{2\mu-1} i!},
\end{aligned}
\]
we obtain
\begin{equation}
\label{ci4}
c_1=-\frac{1}{b}\qquad \text{and }\quad
c_2=\frac{1}{b-1}.
\end{equation}
So finally
\begin{equation}
\label{l3}
\begin{aligned}
L_\mu=&\partial- b^{-1}(\psi_\mu(t,z),zh_1(-z))\partial^{-1}
(\psi_\mu(\tilde t,z),zh_1(-z))+\\[3mm]&\qquad
(b-1)^{-1}(\psi_\mu(t,z),zh_2(-z))\partial^{-1}
(\psi_\mu(\tilde t,z),zh_2(-z))\\[3mm]
=&\partial+(\psi_\mu(t,z),(-)^{\mu+1}\sqrt{-b}e^{-z})\partial^{-1}
(\psi_\mu(\tilde
t,z),(-)^{\mu+1}\sqrt{-b}e^{-z})+\\[3mm]&
\qquad\left(\psi_\mu(t,z),\sqrt{{b-1}}\right)\partial^{-1}
\left(\psi_\mu(\tilde t,z),\sqrt{{b-1}}\right).
\end{aligned}
\end{equation}
We have added the term $(-)^{\mu+1}$  here,
in order to get rid of this term later on in this section.
Note that (see (\ref{tauw'1}))
\[
\begin{aligned}
(\psi_\mu( t,z),(-)^{\mu+1}\sqrt{-b}e^{-z})=&\frac{1}{\sqrt{-b}}
\frac{\tau_{W'}( t)}{\tau_{\mu}(t)}\\[3mm]
=&\frac{1}{\sqrt{-b}}\frac{W\left(
R_0^\mu(t),R_1^\mu(t),R_2^\mu(t),\ldots, R_\mu^\mu(t)\right)}
{W\left(
R_1^\mu(t),R_2^\mu(t),\ldots, R_\mu^\mu(t)\right)},
\end{aligned}
\]
\[
\begin{aligned}
\left(\psi_\mu( t,z),\sqrt{{b-1}}\right)=&\sqrt{{b-1}}
\frac{\tau_{W''}( t)}{\tau_{\mu}( t)}
=&\sqrt{{b-1}}\frac{W\left(
R_1^\mu(t),R_2^\mu(t),\ldots, R_\mu^\mu(t),S_{\mu-1}(t)\right)}
{W\left(
R_1^\mu(t),R_2^\mu(t),\ldots, R_\mu^\mu(t)\right)}.
\end{aligned}
\]
Using (\ref{ci1} ), (\ref{tauww}) and (\ref{ci4}) we find that also
\[
\begin{aligned}
(\psi_\mu(\tilde t,z),(-)^{\mu+1}\sqrt{-b}e^{-z})=&{\sqrt{-b}}
\frac{\tau_{W_\mu+\mathbb{C}zh_1(z)}( t)}{\tau_{\mu}(t)}\\[3mm]
=&{\sqrt{-b}}\frac{W\left(
R^\mu_2(t),R^\mu_3(t), \ldots, R^\mu_\mu (t)\right)}
{W\left(
R_1^\mu(t),R_2^\mu(t),\ldots, R_\mu^\mu(t)\right)},
\end{aligned}
\]
\[
\begin{aligned}
\left(\psi_\mu(\tilde t,z),\sqrt{{b-1}}\right)=&(-)^{\mu-1}\sqrt{{b-1}}
\frac{\tau_{W_\mu+\mathbb{C}zh_2(z)}( t)}{\tau_{\mu}( t)}\\[3mm]
=&(-)^{\mu-1}\sqrt{{b-1}}\frac{W\left(
R^\mu_2(t)-R^\mu_1(t),R^\mu_3(t)-\frac{R^\mu_2(t)}{2},
\ldots, R^\mu_\mu (t)-\frac{R^\mu_{\mu-1}(t)}{\mu-1}\right)}
{W\left(
R_1^\mu(t),R_2^\mu(t),\ldots, R_\mu^\mu(t)\right)}.
\end{aligned}
\]
We thus obtain in this way that
\[
\beta_{12}(s)= \left(\psi_\mu(\tilde s,z),\sqrt{{b-1}}\right)\qquad\mbox{with }
\alpha=\sqrt{{b-1}}
\]
and
\[
\beta_{32}(s)=(\psi_\mu(\tilde s,z),(-)^{\mu+1}\sqrt{-b}e^{-z})\qquad\mbox{with
} a=(-)^{\mu+1}\sqrt{{-b}}.
\]
To obtain $\beta_{13}(s)$, we calculate the so-called squared eigenfunction
potential
\[
\partial^{-1}
\left(\psi_\mu(\tilde t,z),\sqrt{{b-1}}\right)
(\psi_\mu( t,z),(-)^{\mu+1}\sqrt{-b}e^{-z})
\]
of
$\left(\psi_\mu(\tilde t,z),\sqrt{{b-1}}\right)$ and
$(\psi_\mu( t,z),(-)^{\mu+1}\sqrt{-b}e^{-z})$.
Let
\[
w(z)=b_1r_1(-z)+
\sum_{i=1}^{\mu-1}b_{i+1}\left(r_{i+1}(-z)-\frac{ri(-z)}{i}\right)
+\sum_{j>\mu} b_jz^j,
\]
be an arbitrary element of $W_\mu^\perp$, then
\[
\begin{aligned}
&\partial^{-1}
\left(\psi_\mu( \tilde t,z),\sqrt{{b-1}}\right)
(\psi_\mu( t,z),(-)^{\mu+1}\sqrt{-b}e^{-z})\\[3mm]
=&\partial^{-1}\left(\psi_\mu( \tilde t,z),\sqrt{{b-1}}\right)
(\psi_\mu( t,z),(-)^{\mu+1}\sqrt{-b}(e^{-z}+w(z)))\\[3mm]
=&\left(\psi_\mu( \tilde t,z),\sqrt{{b-1}}\right)
\left(\psi_\mu( \tilde t,z),\sqrt{{b-1}}\right)^{-1}
\partial^{-1}\left(\psi_\mu( \tilde t,z),\sqrt{{b-1}}\right)
\times\\[3mm]
&\qquad\qquad\qquad\qquad\qquad\qquad
\left(\psi_\mu( t,z),(-)^{\mu+1}\sqrt{-b}(e^{-z}+w(z))\right)\\[3mm]
=&\left(\psi_\mu( \tilde t,z),\sqrt{{b-1}}\right)
\left(\psi_{W_\mu+\mathbb{C}zh_2(z)}(t,z),
(-)^{\mu+1}\sqrt{-b}(e^{-z}+w(z))\right)\\[3mm]
=&\left(\psi_\mu( \tilde t,z),\sqrt{{b-1}}\right)
\left(\psi_{W_\mu+\mathbb{C}zh_2(z)}(t,z),
(-)^{\mu+1}\sqrt{-b}\left(\frac{r_0(-z)}{b}+b_1r_1(-z)\right)\right).
\end{aligned}
\]
Using (\ref{ci1} ) and (\ref{ci4}) we find that
\begin{equation}
\label{sep3}
\partial^{-1}
\left(\psi_\mu( \tilde t,z),\sqrt{{b-1}}\right)
(\psi_\mu( t,z),(-)^{\mu+1}\sqrt{-b}e^{-z})=
-\sqrt{\frac{b-1}{-b}}\frac{\tau_{W'''}(t)}{\tau_\mu(t)},
\end{equation}
where
\begin{equation}
\label{tauW'''}
\tau_{W'''}(t)=
W\left( R^\mu_2(t)-\frac{R^\mu_1(t)}{1},R^\mu_3(t)-\frac{R^\mu_2(t)}{2},\ldots,
R^\mu_\mu(t)-\frac{R^\mu_{\mu-1}(t)}{\mu-1}, R^\mu_0(t)+bb_1R^\mu_1(t)\right)
{}.
\end{equation}
Now comparing (\ref{bt}), (\ref{tT}), (\ref{sep3}) and  (\ref{tauW'''}) we see
that
\[
b_1=0.
\]
For this $b_1=0$, the tau-function
$\tau_{W'''}$ corresponds to the following point in the Grassmannian:
\begin{equation}
\label{w'''}
W'''=\{ f(z)\in H_{-\mu}|\,(f(z),r_0(-z))=0\ \text{and }
\left(f(z),r_i(-z)-\frac{r_{i-1}(-z)}{i-1}\right)=0\
\text{for } i=2,3\ldots,\mu \}.
\end{equation}
Hence, using the fact that $\alpha=\sqrt{b-1}$ and $a=(-)^{\mu+1} \sqrt{-b}$
one finds that
\[
\beta_{13}(s)=\partial^{-1}
\left(\psi_\mu( \tilde s,z),\sqrt{{b-1}}\right)
(\psi_\mu( s,z),(-)^{\mu+1}\sqrt{-b}e^{-z}).
\]

We now  calculate the squared eigenfunction potential  in a different way.
Let
\[
w(z)=\sum_{i=1}^\mu a_ir_i(-z)+\sum_{j>\mu} a_jz^j,
\]
be an arbitrary element of $W_\mu^\perp$, a straightforward calculation shows
that
\[
\begin{aligned}
&\partial^{-1}(\psi_\mu(\tilde t,z),(-)^{\mu+1}\sqrt{-b}e^{-z})
\left(\psi_\mu(t,z),\sqrt{{b-1}}\right)\\[3mm]
=&\partial^{-1}(\psi_\mu(\tilde t,z),(-)^{\mu+1}\sqrt{-b}e^{-z})
\left(\psi_\mu(t,z),\sqrt{{b-1}}(1+w(z))\right)\\[3mm]
=&(\psi_\mu(\tilde t,z),(-)^{\mu+1}\sqrt{-b}e^{-z})
\left(\psi_{W_\mu+\mathbb{C}zh_1(z)}(t,z),\sqrt{{b-1}}(1+a_1r_1(-z))\right).
\end{aligned}
\]
Using (\ref{ci1} ) and (\ref{ci4}) we find that
\begin{equation}
\label{sep4}
\begin{aligned}
\partial^{-1}(\psi_\mu(\tilde t,z),(-)^{\mu+1}\sqrt{-b}e^{-z})
\left(\psi_\mu(t,z),\sqrt{{b-1}}\right)
=&\sqrt{\frac{b-1}{-b}}\frac{\tau_{W'}(\tilde t)\tau_{W''''}( t)}
{\tau_\mu(\tilde t)\tau_{W_\mu+\mathbb{C}zh_1(z)}(t)}\\[3mm]
=&\sqrt{-b({b-1})}\frac{\tau_{W''''}( t)}{\tau_\mu( t)},
\end{aligned}
\end{equation}
where
\begin{equation}
\label{tauw''''}
\tau_{W''''}(
t)=W\left(R_2^\mu(t),R_3^\mu(t),\ldots,R_\mu^\mu(t),S_{\mu-1}(t)+a_1
R_1^\mu(t)\right).
\end{equation}
This is the tau-function corresponding to the following point of the
Grassmannian
\begin{equation}
\label{W''''}
W''''=\{ f(z)\in H_{-\mu}|\,(f(z),1+a_1r_1(-z))=0\ \text{and }
(f(z),r_i(-z))=0\
\text{for } i=2,3\ldots,\mu \}.
\end{equation}
Hence
\begin{equation}
\label{sep5}
\begin{aligned}
\partial^{-1}(\psi_\mu(\tilde t,z),&(-)^{\mu+1}\sqrt{-b}e^{-z})
\left(\psi_\mu(t,z),\sqrt{{b-1}}\right)\\[3mm]
=&\sqrt{-b({b-1})}\left(
\frac{W\left(
R_2^\mu(t),R_3^\mu(t),\ldots, R_\mu^\mu(t),S_{\mu-1}(t)\right)}
{W\left(
R_1^\mu(t),R_2^\mu(t),\ldots, R_\mu^\mu(t)\right)}-(-)^\mu a_1\right).
\end{aligned}
\end{equation}
It is not clear yet what the value of $a_1$ one should take.

We now put all $t_i=0$ for $i>1$, and write $f(x)$ for $f(x,0,0,\ldots)$.
Comparing (\ref{sep3}) and (\ref{sep4}), we see that
\begin{equation}
\label{btt}
b\tau_{W''''}(x)=\tau_{W'''}(x).
\end{equation}
To calculate $a_1$ we substitute $x=0$. We find that
\begin{equation}
\label{tm0}
\begin{aligned}
\tau_\mu(0)=\tau_{W_\mu}(0)=&
\det\left(
\left(  z^{i-\mu-1},r_{j}(-z)  \right)
\right)_{1\le i,j\le\mu}\\[2mm]
=&
\det\left(
\frac{(-)^{\mu-i}b}{(\mu+j-i)!}
\right)_{1\le i,j\le\mu}\\[2mm]
=&(-)^{\frac{\mu(\mu-1)}{2}}b^{\mu}
\det\left(
\frac{1}{(\mu+j-i)!}
\right)_{1\le i,j\le\mu}\\[2mm]
=&(-)^{\frac{\mu(\mu-1)}{2}}b^{\mu}\mu
S_{\mu,\mu,\ldots,\mu}^{(\mu)}(1,0,0,\ldots)\\[2mm]
=&(-)^{\frac{\mu(\mu-1)}{2}}b^{\mu}\mu
\frac{\prod_{i=1}^{\mu-1} (i!)^2}{\prod_{i=1}^{2\mu-1} i!}.
\end{aligned}
\end{equation}
In a similar way we find that $\tau_{W'''}(0)=\tau_\mu(0)$ and that
\begin{equation}
\label{tm2}
\tau_{W''''}(0)=
(-)^{\frac{\mu(\mu-1)}{2}}(b^{\mu-1}-(-)^\mu a_1 b^\mu)\mu
\frac{\prod_{i=1}^{\mu-1} (i!)^2}{\prod_{i=1}^{2\mu-1} i!}.
\end{equation}
Comparing all the results (\ref{btt}-\ref{tm2}) we conclude that
\[
a_1=0.
\]
Since we know that $\sum_{i=1}^3 \omega_i^2(x)$ is equal to a constant,
it suffices to calculate this value for $x=0$. We find that
\[
\begin{aligned}
\omega_1(0)=&\beta_{32}(0)=\frac{1}{\sqrt{-b}}
\frac{\tau_{W'}(0)}{\tau_{\mu}(0)}\\[3mm]
=&\frac{1}{\sqrt{-b}}
\frac{(-)^{\frac{\mu(\mu-1)}{2}+1}b^\mu
S^{(\mu+1)}_{\mu-1,\mu-1,\ldots,\mu-1}(1,0,0,\ldots)}
{(-)^{\frac{\mu(\mu-1)}{2}}b^\mu
S^{(\mu)}_{\mu,\mu,\ldots,\mu}(1,0,0,\ldots)}\\[3mm]
=&-\frac{1}{\sqrt{-b}}\mu,
\end{aligned}
\]
\[
\begin{aligned}
\omega_2(0)=&-\beta_{13}(0)=
-\sqrt{-b(b-1)}\frac{\tau_{W''''}(0)}{\tau_{\mu}(0)}\\[3mm]
=&-\sqrt{-b(b-1)}
\frac{(-)^{\frac{\mu(\mu-1)}{2}}b^{\mu-1}
S^{(\mu-1)}_{\mu+1,\mu+1,\ldots,\mu+1}(1,0,0,\ldots)}
{(-)^{\frac{\mu(\mu-1)}{2}}b^{\mu-1}
S^{(\mu)}_{\mu,\mu,\ldots,\mu}(1,0,0,\ldots)}\\[3mm]
=&-\sqrt{-b(b-1)}\frac{\mu}{b}
=\sqrt{\frac{b-1}{-b}}\mu,
\end{aligned}
\]
\[
\omega_3(0)=0 \beta_{12}(0)=0.
\]
Hence,
\[
\sum_{i=1}^3 \omega_i^2(0)=\left(\frac{1}{\sqrt{-b}}\mu\right)^2+
\left(\sqrt{\frac{b-1}{-b}}\mu\right)^2=-\mu^2
\]
and
\begin{equation}
\label{omegakwad}
\sum_{i=1}^3 \omega_i^2(x)=-\mu^2.
\end{equation}
We next calculate $R_i^\mu(x)$:
\[
\begin{aligned}
R_i^\mu(x)=&\sum_{k=0}^{i-1} \frac{(-1)^{k-i}}{k!}
\frac{x^{\mu+i-k-1}}{(\mu+i-k-1)!} +
b\sum_{k=i}^{\mu+i-1} \frac{(-1)^{k-i}}{k!}
\frac{x^{\mu+i-k-1}}{(\mu+i-k-1)!}\\[3mm]
=&
(-)^i\frac{(x-1)^{\mu+i-1}}{(\mu+i-1)!}+(-)^{\mu-1}(b-1)
\sum_{j=0}^{\mu-1}
\frac{(-x)^{j}}{j!(\mu+i-j-1)!}.
\end{aligned}
\]

Combining all the previous results we find:

\begin{theorem}
The expressions
\[
y(x) = x \frac{\pm (x-1) \mu \omega_1 \omega_2 \omega_3 + x \omega_1^2
\omega_2^2+\omega_1^2 \omega_3^2}{(x-1)^2 \omega_2^2 \omega_3^2 +x^2
\omega_1^2 \omega_2^2+\omega_1^2 \omega_3^2 }
\]
and
\[
y(x) = - x \frac{x(\om_1\om_2 \mp \mu\om_3)^2+(\om_1\om_3 \pm \mu\om_2)^2}{
\(\om_3^2+\mu^2+x (\om_2^2+\mu^2)\)^2+4 x \mu^2\om_1^2}
\]
for $\mu=1,2,\ldots$,
with
\begin{equation}
\label{valomega}
\begin{aligned}
\omega_1(x)&=\sqrt{-b}(1-x)\frac{W\left(
R_2^\mu(x),R_3^\mu(x),\ldots, R_\mu^\mu(x)\right)}
{W\left(
R_1^\mu(x),R_2^\mu(x),\ldots, R_\mu^\mu(x)\right)},
\\[3mm]
\omega_2(x)&=-\sqrt{-b({b-1})}
\frac{W\left(
R_2^\mu(x),R_3^\mu(x),\ldots, R_\mu^\mu(x),\frac{x^{\mu-1}}{(\mu-1)!}\right)}
{W\left(
R_1^\mu(x),R_2^\mu(x),\ldots, R_\mu^\mu(x)\right)},
\\[3mm]
\omega_3(x)&=\sqrt{{b-1}}x\frac{W\left(
R_1^\mu(x),R_2^\mu(x),\ldots, R_\mu^\mu(x),\frac{x^{\mu-1}}{(\mu-1)!}\right)}
{W\left(
R_1^\mu(x),R_2^\mu(x),\ldots, R_\mu^\mu(x)\right)}
\end{aligned}
\end{equation}
are rational solutions of  the Painlev\'e VI equation (\ref{eq:pain6})
for the parameters
\[
\begin{aligned}
(\alpha,\beta,\gamma ,\delta )=&
\left(\frac{(1\mp \mu)^2}{2},- \frac{\mu^2}{2},
\frac{\mu^2}{2}, \frac{1-\mu^2}{2}\right),\quad\mbox{respectively}\\[2mm]
(\alpha,\beta,\gamma ,\delta )=&
\left(\frac{(1 \pm 2\mu)^2}{2},0,0,\frac{1}{2}\right)
{}.
\end{aligned}
\]
The  $\omega_i$ separately satisfy the time dependent Euler
top equations (\ref{eq:euta}).
\end{theorem}
The above results are clearly valid for $\mu>1$.
We will now treat the case $\mu=1$ separately. In that case $W_1$
corresponds to the 1-vector 1-constrained KP hierarchy and
\[
\tau_1(t)=R_1^1(t)=-S_1(t)+bS_0(t)=b-x.
\]
We use the same expressions for the $\beta_{ij}(x)$ in terms of the
Wronskian determinants as in the case $\mu>1$, viz.,
\[
\beta_{23}(x)= \frac{1}{\sqrt{-b}}\frac{-b}{b-x},\quad
\beta_{12}(x)=\sqrt{b-1}\frac{1}{b-x},\quad
\beta_{13}(x)=\sqrt{-b(b-1)}\frac{1}{b-x}.
\]
This leads to the $\omega_i$'s (\ref{eq:omeone}) for $\mu^2=1$.
\begin{remark}From the rational solutions (\ref{valomega}) for the time
dependent Euler top
equations  for the values $\mu=1,2,3,\ldots$ one can
recover the expression of the $\omega_i$ in
  the  $u_i$, $i=1,2,3$, by just substituting:
\[
x=\frac{u_2-u_1}{u_3-u_1}
\]
in $V(x)$, i.e., in all $\omega_i(x)$.
Using (\ref{eq:vome}) one finds expressions for the rotation coefficients
$\beta_{ij}(u)$ that satisfy (\ref{betas-comp})-(\ref{betas-deg}).
\end{remark}

Finally we give as an example the explicit $\omega_i$'s for $\mu=3$:
{\scriptsize
\[
\begin{aligned}
\omega_1(x)=&{
\frac{3{\sqrt{-b}}\left( 1 - x \right)
    \left( b^2 - 8b^2x + 18bx^2 + 10b^2x^2 -
      56bx^3 + 70bx^4 - 56bx^5 + 10x^6 +
      18bx^6 - 8x^7 + x^8 \right) }{
    b^3 - 9b^2x + 36b^2x^2 - 84b^2x^3 +
      36bx^4 + 90b^2x^4 - 90bx^5 -
      36b^2x^5 + 84bx^6 - 36bx^7 + 9bx^8 -
      x^9 },}\\[3mm]
\omega_2(x)=& {\frac{-3{\sqrt{-b(b-1)}}
    \left( b^2 - 18bx^2 + 52bx^3 - 60bx^4 +
      24bx^5 + 10x^6 - 12x^7 + 3x^8 \right) }
    {b^3 - 9b^2x + 36b^2x^2 - 84b^2x^3 +
    36bx^4 + 90b^2x^4 - 90bx^5 - 36b^2x^5 +
    84bx^6 - 36bx^7 + 9bx^8 - x^9},}\\[3mm]
\omega_3(x)=& {\frac{3{\sqrt{b-1}}x
    \left( 3b^2 - 12b^2x + 10b^2x^2 +
      24bx^3 - 60bx^4 + 52bx^5 - 18bx^6 +
      x^8 \right) }{b^3 - 9b^2x + 36b^2x^2 -
    84b^2x^3 + 36bx^4 + 90b^2x^4 - 90bx^5 -
    36b^2x^5 + 84bx^6 - 36bx^7 + 9bx^8 - x^9
    }.}
      \end{aligned}
\]}

\subsection*{Acknowledgement: \rm JvdL wishes to thank Pierre van Moerbeke for
valuable
conversations.}

\end{document}